\documentstyle[prl,aps,graphicx]{revtex}
\begin{document}
\draft
\title{Boost-rotation symmetric spacetimes -- review}
\def \BE {\begin{equation}}
\def \EE {\end{equation}}
\def \BEAH {\begin{eqnarray*}}
\def \EEAH {\end{eqnarray*}}
\def \BEA {\begin{eqnarray}}
\def \EEA {\end{eqnarray}}
\def \BDM {\begin{displaymath}}
\def \EDM {\end{displaymath}}
\def \t {\tau}
\def \s {\sigma}
\def \bfi {\bar \varphi}
\def \br {\bar \rho}
\def \bz {\bar z}
\def \bt {\bar t}
\def \e {\varepsilon}
\def \g {\gamma}
\def \gab {g_{\a \b }}
\def \gmn {g_{\m \n }}
\def \gma {g_{\m \a }}
\def \gnn {g_{00}}
\def \gnj {g_{01}}
\def \gnd {g_{02}}
\def \gnt {g_{03}}
\def \gjn {g_{10}}
\def \gjj {g_{11}}
\def \gjd {g_{12}}
\def \gjt {g_{13}}
\def \gdn {g_{20}}
\def \gdj {g_{21}}
\def \gdd {g_{22}}
\def \gdt {g_{23}}
\def \gtn {g_{30}}
\def \gtj {g_{31}}
\def \gtd {g_{32}}
\def \gtt {g_{33}}
\def \guab {g^{\a \b }}
\def \guae {g^{\a \ep }}
\def \gumn {g^{\m \n }}
\def \gunn {g^{00}}
\def \gunj {g^{01}}
\def \gund {g^{02}}
\def \gunt {g^{03}}
\def \gujn {g^{10}}
\def \gujj {g^{11}}
\def \gujd {g^{12}}
\def \gujt {g^{13}}
\def \gudn {g^{20}}
\def \gudj {g^{21}}
\def \gudd {g^{22}}
\def \gudt {g^{23}}
\def \gutn {g^{30}}
\def \gutj {g^{31}}
\def \gutd {g^{32}}
\def \gutt {g^{33}}
\def \gabu {\gab,_u}
\def \gmnu {\gmn,_u}
\def \gnnu {\gnn,_u}
\def \gnju {\gnj,_u}
\def \gndu {\gnd,_u}
\def \gntu {\gnt,_u}
\def \gjnu {\gjn,_u}
\def \gjju {\gjj,_u}
\def \gjdu {\gjd,_u}
\def \gjtu {\gjt,_u}
\def \gdnu {\gdn,_u}
\def \gdju {\gdj,_u}
\def \gddu {\gdd,_u}
\def \gdtu {\gdt,_u}
\def \gtnu {\gtn,_u}
\def \gtju {\gtj,_u}
\def \gtdu {\gtd,_u}
\def \gttu {\gtt,_u}
\def \guabu {\guab,_u}
\def \gumnu {\gumn,_u}
\def \gunnu {\gunn,_u}
\def \gunju {\gunj,_u}
\def \gundu {\gund,_u}
\def \guntu {\gunt,_u}
\def \gujnu {\gujn,_u}
\def \gujju {\gujj,_u}
\def \gujdu {\gujd,_u}
\def \gujtu {\gujt,_u}
\def \gudnu {\gudn,_u}
\def \gudju {\gudj,_u}
\def \guddu {\gudd,_u}
\def \gudtu {\gudt,_u}
\def \gutnu {\gutn,_u}
\def \gutju {\gutj,_u}
\def \gutdu {\gutd,_u}
\def \guttu {\gutt,_u}
\def \gabr {\gab,_r}
\def \gmnr {\gmn,_r}
\def \gnnr {\gnn,_r}
\def \gnjr {\gnj,_r}
\def \gndr {\gnd,_r}
\def \gntr {\gnt,_r}
\def \gjnr {\gjn,_r}
\def \gjjr {\gjj,_r}
\def \gjdr {\gjd,_r}
\def \gjtr {\gjt,_r}
\def \gdnr {\gdn,_r}
\def \gdjr {\gdj,_r}
\def \gddr {\gdd,_r}
\def \gdtr {\gdt,_r}
\def \gtnr {\gtn,_r}
\def \gtjr {\gtj,_r}
\def \gtdr {\gtd,_r}
\def \gttr {\gtt,_r}
\def \guabr {\guab,_r}
\def \gumnr {\gumn,_r}
\def \gunnr {\gunn,_r}
\def \gunjr {\gunj,_r}
\def \gundr {\gund,_r}
\def \guntr {\gunt,_r}
\def \gujnr {\gujn,_r}
\def \gujjr {\gujj,_r}
\def \gujdr {\gujd,_r}
\def \gujtr {\gujt,_r}
\def \gudnr {\gudn,_r}
\def \gudjr {\gudj,_r}
\def \guddr {\gudd,_r}
\def \gudtr {\gudt,_r}
\def \gutnr {\gutn,_r}
\def \gutjr {\gutj,_r}
\def \gutdr {\gutd,_r}
\def \guttr {\gutt,_r}
\def \gabt {\gab,_\th }
\def \gmnt {\gmn,_\th }
\def \gnnt {\gnn,_\th }
\def \gnjt {\gnj,_\th }
\def \gndt {\gnd,_\th }
\def \gntt {\gnt,_\th }
\def \gjnt {\gjn,_\th }
\def \gjjt {\gjj,_\th }
\def \gjdt {\gjd,_\th }
\def \gjtt {\gjt,_\th }
\def \gdnt {\gdn,_\th }
\def \gdjt {\gdj,_\th }
\def \gddt {\gdd,_\th }
\def \gdtt {\gdt,_\th }
\def \gtnt {\gtn,_\th }
\def \gtjt {\gtj,_\th }
\def \gtdt {\gtd,_\th }
\def \gttt {\gtt,_\th }
\def \guabt {\guab,_\th }
\def \gumnt {\gumn,_\th }
\def \gunnt {\gunn,_\th }
\def \gunjt {\gunj,_\th }
\def \gundt {\gund,_\th }
\def \guntt {\gunt,_\th }
\def \gujnt {\gujn,_\th }
\def \gujjt {\gujj,_\th }
\def \gujdt {\gujd,_\th }
\def \gujtt {\gujt,_\th }
\def \gudnt {\gudn,_\th }
\def \gudjt {\gudj,_\th }
\def \guddt {\gudd,_\th }
\def \gudtt {\gudt,_\th }
\def \gutnt {\gutn,_\th }
\def \gutjt {\gutj,_\th }
\def \gutdt {\gutd,_\th }
\def \guttt {\gutt,_\th }
\def \mm {\mbox{\quad }}
\def \mv {\mbox{\qquad }}
\def \msip {\rightarrow}
\def \vsip {\longrightarrow}
\def \Fab {F_{\a \b }}
\def \Fmn {F_{\m \n }}
\def \Fgb {F_{\g \b }}
\def \Fag {F_{\a \g }}
\def \Fnn {F_{00}}
\def \Fnj {F_{01}}
\def \Fnd {F_{02}}
\def \Fnt {F_{03}}
\def \Fjn {F_{10}}
\def \Fjj {F_{11}}
\def \Fjd {F_{12}}
\def \Fjt {F_{13}}
\def \Fdn {F_{20}}
\def \Fdj {F_{21}}
\def \Fdd {F_{22}}
\def \Fdt {F_{23}}
\def \Ftn {F_{30}}
\def \Ftj {F_{31}}
\def \Ftd {F_{32}}
\def \Ftt {F_{33}}
\def \Fuab {F^{\a \b }}
\def \Fumn {F^{\m \n }}
\def \Funn {F^{00}}
\def \Funj {F^{01}}
\def \Fund {F^{02}}
\def \Funt {F^{03}}
\def \Fujn {F^{10}}
\def \Fujj {F^{11}}
\def \Fujd {F^{12}}
\def \Fujt {F^{13}}
\def \Fudn {F^{20}}
\def \Fudj {F^{21}}
\def \Fudd {F^{22}}
\def \Fudt {F^{23}}
\def \Futn {F^{30}}
\def \Futj {F^{31}}
\def \Futd {F^{32}}
\def \Futt {F^{33}}
\def \Fabu {F_{\a \b },_u}
\def \Fmnu {F_{\m \n },_u}
\def \Fnnu {F_{00},_u}
\def \Fnju {F_{01},_u}
\def \Fndu {F_{02},_u}
\def \Fntu {F_{03},_u}
\def \Fjnu {F_{10},_u}
\def \Fjju {F_{11},_u}
\def \Fjdu {F_{12},_u}
\def \Fjtu {F_{13},_u}
\def \Fdnu {F_{20},_u}
\def \Fdju {F_{21},_u}
\def \Fddu {F_{22},_u}
\def \Fdtu {F_{23},_u}
\def \Ftnu {F_{30},_u}
\def \Ftju {F_{31},_u}
\def \Ftdu {F_{32},_u}
\def \Fttu {F_{33},_u}
\def \Fuabu {F^{\a \b },_u}
\def \Fumnu {F^{\m \n },_u}
\def \Funnu {F^{00},_u}
\def \Funju {F^{01},_u}
\def \Fundu {F^{02},_u}
\def \Funtu {F^{03},_u}
\def \Fujnu {F^{10},_u}
\def \Fujju {F^{11},_u}
\def \Fujdu {F^{12},_u}
\def \Fujtu {F^{13},_u}
\def \Fudnu {F^{20},_u}
\def \Fudju {F^{21},_u}
\def \Fuddu {F^{22},_u}
\def \Fudtu {F^{23},_u}
\def \Futnu {F^{30},_u}
\def \Futju {F^{31},_u}
\def \Futdu {F^{32},_u}
\def \Futtu {F^{33},_u}
\def \Fabr {F_{\a \b },_r}
\def \Fmnr {F_{\m \n },_r}
\def \Fnnr {F_{00},_r}
\def \Fnjr {F_{01},_r}
\def \Fndr {F_{02},_r}
\def \Fntr {F_{03},_r}
\def \Fjnr {F_{10},_r}
\def \Fjjr {F_{11},_r}
\def \Fjdr {F_{12},_r}
\def \Fjtr {F_{13},_r}
\def \Fdnr {F_{20},_r}
\def \Fdjr {F_{21},_r}
\def \Fddr {F_{22},_r}
\def \Fdtr {F_{23},_r}
\def \Ftnr {F_{30},_r}
\def \Ftjr {F_{31},_r}
\def \Ftdr {F_{32},_r}
\def \Fttr {F_{33},_r}
\def \Fuabr {F^{\a \b },_r}
\def \Fumnr {F^{\m \n },_r}
\def \Funnr {F^{00},_r}
\def \Funjr {F^{01},_r}
\def \Fundr {F^{02},_r}
\def \Funtr {F^{03},_r}
\def \Fujnr {F^{10},_r}
\def \Fujjr {F^{11},_r}
\def \Fujdr {F^{12},_r}
\def \Fujtr {F^{13},_r}
\def \Fudnr {F^{20},_r}
\def \Fudjr {F^{21},_r}
\def \Fuddr {F^{22},_r}
\def \Fudtr {F^{23},_r}
\def \Futnr {F^{30},_r}
\def \Futjr {F^{31},_r}
\def \Futdr {F^{32},_r}
\def \Futtr {F^{33},_r}
\def \Fabt {F_{\a \b },_\th }
\def \Fmnt {F_{\m \n },_\th }
\def \Fnnt {F_{00},_\th }
\def \Fnjt {F_{01},_\th }
\def \Fndt {F_{02},_\th }
\def \Fntt {F_{03},_\th }
\def \Fjnt {F_{10},_\th }
\def \Fjjt {F_{11},_\th }
\def \Fjdt {F_{12},_\th }
\def \Fjtt {F_{13},_\th }
\def \Fdnt {F_{20},_\th }
\def \Fdjt {F_{21},_\th }
\def \Fddt {F_{22},_\th }
\def \Fdtt {F_{23},_\th }
\def \Ftnt {F_{30},_\th }
\def \Ftjt {F_{31},_\th }
\def \Ftdt {F_{32},_\th }
\def \Fttt {F_{33},_\th }
\def \Fuabt {F^{\a \b },_\th }
\def \Fumnt {F^{\m \n },_\th }
\def \Funnt {F^{00},_\th }
\def \Funjt {F^{01},_\th }
\def \Fundt {F^{02},_\th }
\def \Funtt {F^{03},_\th }
\def \Fujnt {F^{10},_\th }
\def \Fujjt {F^{11},_\th }
\def \Fujdt {F^{12},_\th }
\def \Fujtt {F^{13},_\th }
\def \Fudnt {F^{20},_\th }
\def \Fudjt {F^{21},_\th }
\def \Fuddt {F^{22},_\th }
\def \Fudtt {F^{23},_\th }
\def \Futnt {F^{30},_\th }
\def \Futjt {F^{31},_\th }
\def \Futdt {F^{32},_\th }
\def \Futtt {F^{33},_\th }
\def \Emn {E_{\m\n}}
\def \Eab {E_{\a \b }}
\def \Egb {E_{\g \b }}
\def \Eag {E_{\a \g }}
\def \Enn {E_{00}}
\def \Enj {E_{01}}
\def \End {E_{02}}
\def \Ent {E_{03}}
\def \Ejn {E_{10}}
\def \Ejj {E_{11}}
\def \Ejd {E_{12}}
\def \Ejt {E_{13}}
\def \Edn {E_{20}}
\def \Edj {E_{21}}
\def \Edd {E_{22}}
\def \Edt {E_{23}}
\def \Etn {E_{30}}
\def \Etj {E_{31}}
\def \Etd {E_{32}}
\def \Ett {E_{33}}
\def \Euab {E^{\a \b }}
\def \Eumn {E^{\m \n }}
\def \Eunn {E^{00}}
\def \Eunj {E^{01}}
\def \Eund {E^{02}}
\def \Eunt {E^{03}}
\def \Eujn {E^{10}}
\def \Eujj {E^{11}}
\def \Eujd {E^{12}}
\def \Eujt {E^{13}}
\def \Eudn {E^{20}}
\def \Eudj {E^{21}}
\def \Eudd {E^{22}}
\def \Eudt {E^{23}}
\def \Eutn {E^{30}}
\def \Eutj {E^{31}}
\def \Eutd {E^{32}}
\def \Eutt {E^{33}}
\def \Eabu {E_{\a \b },_u}
\def \Emnu {E_{\m \n },_u}
\def \Ennu {E_{00},_u}
\def \Enju {E_{01},_u}
\def \Endu {E_{02},_u}
\def \Entu {E_{03},_u}
\def \Ejnu {E_{10},_u}
\def \Ejju {E_{11},_u}
\def \Ejdu {E_{12},_u}
\def \Ejtu {E_{13},_u}
\def \Ednu {E_{20},_u}
\def \Edju {E_{21},_u}
\def \Eddu {E_{22},_u}
\def \Edtu {E_{23},_u}
\def \Etnu {E_{30},_u}
\def \Etju {E_{31},_u}
\def \Etdu {E_{32},_u}
\def \Ettu {E_{33},_u}
\def \Euabu {E^{\a \b },_u}
\def \Eumnu {E^{\m \n },_u}
\def \Eunnu {E^{00},_u}
\def \Eunju {E^{01},_u}
\def \Eundu {E^{02},_u}
\def \Euntu {E^{03},_u}
\def \Eujnu {E^{10},_u}
\def \Eujju {E^{11},_u}
\def \Eujdu {E^{12},_u}
\def \Eujtu {E^{13},_u}
\def \Eudnu {E^{20},_u}
\def \Eudju {E^{21},_u}
\def \Euddu {E^{22},_u}
\def \Eudtu {E^{23},_u}
\def \Eutnu {E^{30},_u}
\def \Eutju {E^{31},_u}
\def \Eutdu {E^{32},_u}
\def \Euttu {E^{33},_u}
\def \Eabr {E_{\a \b },_r}
\def \Emnr {E_{\m \n },_r}
\def \Ennr {E_{00},_r}
\def \Enjr {E_{01},_r}
\def \Endr {E_{02},_r}
\def \Entr {E_{03},_r}
\def \Ejnr {E_{10},_r}
\def \Ejjr {E_{11},_r}
\def \Ejdr {E_{12},_r}
\def \Ejtr {E_{13},_r}
\def \Ednr {E_{20},_r}
\def \Edjr {E_{21},_r}
\def \Eddr {E_{22},_r}
\def \Edtr {E_{23},_r}
\def \Etnr {E_{30},_r}
\def \Etjr {E_{31},_r}
\def \Etdr {E_{32},_r}
\def \Ettr {E_{33},_r}
\def \Euabr {E^{\a \b },_r}
\def \Eumnr {E^{\m \n },_r}
\def \Eunnr {E^{00},_r}
\def \Eunjr {E^{01},_r}
\def \Eundr {E^{02},_r}
\def \Euntr {E^{03},_r}
\def \Eujnr {E^{10},_r}
\def \Eujjr {E^{11},_r}
\def \Eujdr {E^{12},_r}
\def \Eujtr {E^{13},_r}
\def \Eudnr {E^{20},_r}
\def \Eudjr {E^{21},_r}
\def \Euddr {E^{22},_r}
\def \Eudtr {E^{23},_r}
\def \Eutnr {E^{30},_r}
\def \Eutjr {E^{31},_r}
\def \Eutdr {E^{32},_r}
\def \Euttr {E^{33},_r}
\def \Eabt {E_{\a \b },_\th }
\def \Emnt {E_{\m \n },_\th }
\def \Ennt {E_{00},_\th }
\def \Enjt {E_{01},_\th }
\def \Endt {E_{02},_\th }
\def \Entt {E_{03},_\th }
\def \Ejnt {E_{10},_\th }
\def \Ejjt {E_{11},_\th }
\def \Ejdt {E_{12},_\th }
\def \Ejtt {E_{13},_\th }
\def \Ednt {E_{20},_\th }
\def \Edjt {E_{21},_\th }
\def \Eddt {E_{22},_\th }
\def \Edtt {E_{23},_\th }
\def \Etnt {E_{30},_\th }
\def \Etjt {E_{31},_\th }
\def \Etdt {E_{32},_\th }
\def \Ettt {E_{33},_\th }
\def \Euabt {E^{\a \b },_\th }
\def \Eumnt {E^{\m \n },_\th }
\def \Eunnt {E^{00},_\th }
\def \Eunjt {E^{01},_\th }
\def \Eundt {E^{02},_\th }
\def \Euntt {E^{03},_\th }
\def \Eujnt {E^{10},_\th }
\def \Eujjt {E^{11},_\th }
\def \Eujdt {E^{12},_\th }
\def \Eujtt {E^{13},_\th }
\def \Eudnt {E^{20},_\th }
\def \Eudjt {E^{21},_\th }
\def \Euddt {E^{22},_\th }
\def \Eudtt {E^{23},_\th }
\def \Eutnt {E^{30},_\th }
\def \Eutjt {E^{31},_\th }
\def \Eutdt {E^{32},_\th }
\def \Euttt {E^{33},_\th }
\def \Tmn {T_{\m\n}}
\def \Tab {T_{\a \b }}
\def \Tgb {T_{\g \b }}
\def \Tag {T_{\a \g }}
\def \Tnn {T_{00}}
\def \Tnj {T_{01}}
\def \Tnd {T_{02}}
\def \Tnt {T_{03}}
\def \Tjn {T_{10}}
\def \Tjj {T_{11}}
\def \Tjd {T_{12}}
\def \Tjt {T_{13}}
\def \Tdn {T_{20}}
\def \Tdj {T_{21}}
\def \Tdd {T_{22}}
\def \Tdt {T_{23}}
\def \Ttn {T_{30}}
\def \Ttj {T_{31}}
\def \Ttd {T_{32}}
\def \Ttt {T_{33}}
\def \Tuab {T^{\a \b }}
\def \Tumn {T^{\m \n }}
\def \Tunn {T^{00}}
\def \Tunj {T^{01}}
\def \Tund {T^{02}}
\def \Tunt {T^{03}}
\def \Tujn {T^{10}}
\def \Tujj {T^{11}}
\def \Tujd {T^{12}}
\def \Tujt {T^{13}}
\def \Tudn {T^{20}}
\def \Tudj {T^{21}}
\def \Tudd {T^{22}}
\def \Tudt {T^{23}}
\def \Tutn {T^{30}}
\def \Tutj {T^{31}}
\def \Tutd {T^{32}}
\def \Tutt {T^{33}}
\def \Tabu {T_{\a \b },_u}
\def \Tmnu {T_{\m \n },_u}
\def \Tnnu {T_{00},_u}
\def \Tnju {T_{01},_u}
\def \Tndu {T_{02},_u}
\def \Tntu {T_{03},_u}
\def \Tjnu {T_{10},_u}
\def \Tjju {T_{11},_u}
\def \Tjdu {T_{12},_u}
\def \Tjtu {T_{13},_u}
\def \Tdnu {T_{20},_u}
\def \Tdju {T_{21},_u}
\def \Tddu {T_{22},_u}
\def \Tdtu {T_{23},_u}
\def \Ttnu {T_{30},_u}
\def \Ttju {T_{31},_u}
\def \Ttdu {T_{32},_u}
\def \Tttu {T_{33},_u}
\def \Tuabu {T^{\a \b },_u}
\def \Tumnu {T^{\m \n },_u}
\def \Tunnu {T^{00},_u}
\def \Tunju {T^{01},_u}
\def \Tundu {T^{02},_u}
\def \Tuntu {T^{03},_u}
\def \Tujnu {T^{10},_u}
\def \Tujju {T^{11},_u}
\def \Tujdu {T^{12},_u}
\def \Tujtu {T^{13},_u}
\def \Tudnu {T^{20},_u}
\def \Tudju {T^{21},_u}
\def \Tuddu {T^{22},_u}
\def \Tudtu {T^{23},_u}
\def \Tutnu {T^{30},_u}
\def \Tutju {T^{31},_u}
\def \Tutdu {T^{32},_u}
\def \Tuttu {T^{33},_u}
\def \Tabr {T_{\a \b },_r}
\def \Tmnr {T_{\m \n },_r}
\def \Tnnr {T_{00},_r}
\def \Tnjr {T_{01},_r}
\def \Tndr {T_{02},_r}
\def \Tntr {T_{03},_r}
\def \Tjnr {T_{10},_r}
\def \Tjjr {T_{11},_r}
\def \Tjdr {T_{12},_r}
\def \Tjtr {T_{13},_r}
\def \Tdnr {T_{20},_r}
\def \Tdjr {T_{21},_r}
\def \Tddr {T_{22},_r}
\def \Tdtr {T_{23},_r}
\def \Ttnr {T_{30},_r}
\def \Ttjr {T_{31},_r}
\def \Ttdr {T_{32},_r}
\def \Tttr {T_{33},_r}
\def \Tuabr {T^{\a \b },_r}
\def \Tumnr {T^{\m \n },_r}
\def \Tunnr {T^{00},_r}
\def \Tunjr {T^{01},_r}
\def \Tundr {T^{02},_r}
\def \Tuntr {T^{03},_r}
\def \Tujnr {T^{10},_r}
\def \Tujjr {T^{11},_r}
\def \Tujdr {T^{12},_r}
\def \Tujtr {T^{13},_r}
\def \Tudnr {T^{20},_r}
\def \Tudjr {T^{21},_r}
\def \Tuddr {T^{22},_r}
\def \Tudtr {T^{23},_r}
\def \Tutnr {T^{30},_r}
\def \Tutjr {T^{31},_r}
\def \Tutdr {T^{32},_r}
\def \Tuttr {T^{33},_r}
\def \Tabt {T_{\a \b },_\th }
\def \Tmnt {T_{\m \n },_\th }
\def \Tnnt {T_{00},_\th }
\def \Tnjt {T_{01},_\th }
\def \Tndt {T_{02},_\th }
\def \Tntt {T_{03},_\th }
\def \Tjnt {T_{10},_\th }
\def \Tjjt {T_{11},_\th }
\def \Tjdt {T_{12},_\th }
\def \Tjtt {T_{13},_\th }
\def \Tdnt {T_{20},_\th }
\def \Tdjt {T_{21},_\th }
\def \Tddt {T_{22},_\th }
\def \Tdtt {T_{23},_\th }
\def \Ttnt {T_{30},_\th }
\def \Ttjt {T_{31},_\th }
\def \Ttdt {T_{32},_\th }
\def \Tttt {T_{33},_\th }
\def \Tuabt {T^{\a \b },_\th }
\def \Tumnt {T^{\m \n },_\th }
\def \Tunnt {T^{00},_\th }
\def \Tunjt {T^{01},_\th }
\def \Tundt {T^{02},_\th }
\def \Tuntt {T^{03},_\th }
\def \Tujnt {T^{10},_\th }
\def \Tujjt {T^{11},_\th }
\def \Tujdt {T^{12},_\th }
\def \Tujtt {T^{13},_\th }
\def \Tudnt {T^{20},_\th }
\def \Tudjt {T^{21},_\th }
\def \Tuddt {T^{22},_\th }
\def \Tudtt {T^{23},_\th }
\def \Tutnt {T^{30},_\th }
\def \Tutjt {T^{31},_\th }
\def \Tutdt {T^{32},_\th }
\def \Tuttt {T^{33},_\th }
\def \a {\alpha}
\def \b {\beta}
\def \g {\gamma}
\def \G {\Gamma}
\def \d {\delta}
\def \eps {\varepsilon}
\def \ep {\epsilon}
\def \e {\eta}
\def \f {\phi}
\def \ffi {\varphi}
\def \j {\iota}
\def \th {\theta}
\def \vth {\vartheta}
\def \k {\kappa}
\def \l {\lambda}
\def \m {\mu}
\def \n {\nu}
\def \x {\xi}
\def \p {\pi}
\def \r {\rho}
\def \s {\sigma}
\def \t {\tau}
\def \ps {\psi}
\def \o {\omega}
\def \z {\zeta}
\def \L {\pounds}
\def \vu {\tilde u}
\def \der {\partial }
\def \nn {\nonumber}
\def \rov {\equiv}
\def \A {{\cal A}}
\def \BB {{\cal B}}
\def \C {{\cal C}}
\def \D {{\cal D}}
\def \E {{\cal E}}
\def \F {{\cal F}}
\def \GG {{\cal G}}
\def \K {{\cal K}}
\def \N {{\cal N}}
\def \L {{\cal L}}
\def \cN {\bar\N}
\def \vN {\tilde\N}
\def \M {{\cal M}}
\def \vM {\tilde \M}
\def \I {{\cal I}}
\def \J {{\cal J}}
\def \R {{\cal R}}
\def \CC {{\rm C}}
\def \U  {{\cal U}}
\def \T  {{\cal T}}
\def \fk  {f^{(k)}}
\def \fmj {f^{(-1)}}
\def \fn  {f^{(0)}}
\def \fj  {f^{(1)}}
\def \fd  {f^{(2)}}
\def \ft  {f^{(3)}}
\def \Ak  {A^{(k)}}
\def \Amj {A^{(-1)}}
\def \An  {A^{(0)}}
\def \Aj  {A^{(1)}}
\def \Ad  {A^{(2)}}
\def \At  {A^{(3)}}
\def \gk  {g^{(k)}}
\def \gmj {g^{(-1)}}
\def \gen  {g^{(0)}}
\def \gej  {g^{(1)}}
\def \ged  {g^{(2)}}
\def \get  {g^{(3)}}
\def \bn  {\b_{0}}
\def \bu  {\b,_{u}}
\def \bj  {\b_{1}}
\def \br  {\b,_{r}}
\def \bd  {\b_{2}}
\def \bth {\b,_{\th}}
\def \bt  {\b_{3}}
\def \bfi  {\b,_{\f}}
\def \bjj   {\b_{11}}
\def \brr   {\b,_{rr}}
\def \bjd   {\b_{12}}
\def \brt   {\b,_{r\th}}
\def \bjt   {\b_{13}}
\def \brf   {\b,_{r\f}}
\def \bdd   {\b_{22}}
\def \bthth {\b,_{\th\th}}
\def \bdt   {\b_{23}}
\def \btf   {\b,_{\th\f}}
\def \btt   {\b_{33}}
\def \bff   {\b,_{\f\f}}
\def \gn  {\g_{0}}
\def \gu  {\g,_{u}}
\def \gj  {\g_{1}}
\def \gr  {\g,_{r}}
\def \gd  {\g_{2}}
\def \gth {\g,_{\th}}
\def \gt  {\g_{3}}
\def \gf  {\g,_{\f}}
\def \gajj  {\g_{11}}
\def \grr   {\g,_{rr}}
\def \gajd  {\g_{12}}
\def \grt   {\g,_{r\th}}
\def \gajt  {\g_{13}}
\def \grf   {\g,_{r\f}}
\def \gadd  {\g_{22}}
\def \gthth {\g,_{\th\th}}
\def \gadt  {\g_{23}}
\def \gtf   {\g,_{\th\f}}
\def \gatt  {\g_{33}}
\def \gff   {\g,_{\f\f}}
\def \dn  {\d_{0}}
\def \du  {\d,_{u}}
\def \dj  {\d_{1}}
\def \dr  {\d,_{r}}
\def \dD  {\d_{2}}
\def \dth {\d,_{\th}}
\def \dt  {\d_{3}}
\def \df  {\d,_{\f}}
\def \djj   {\d_{11}}
\def \drr   {\d,_{rr}}
\def \djd   {\d_{12}}
\def \drt   {\d,_{r\th}}
\def \djt   {\d_{13}}
\def \drf   {\d,_{r\f}}
\def \ddd   {\d_{22}}
\def \dthth {\d,_{\th\th}}
\def \ddt   {\d_{23}}
\def \dtf   {\d,_{\th\f}}
\def \dtt   {\d_{33}}
\def \dff   {\d,_{\f\f}}
\def \Vn   {V_{0}}
\def \Vu   {V,_{u}}
\def \Vj   {V_{1}}
\def \Vr   {V,_{r}}
\def \Vd   {V_{2}}
\def \Vth  {V,_{\th}}
\def \Vt   {V_{3}}
\def \Vf   {V,_{\f}}
\def \Vjj   {V_{11}}
\def \Vrr   {V,_{rr}}
\def \Vjd   {V_{12}}
\def \Vrt   {V,_{r\th}}
\def \Vjt   {V_{13}}
\def \Vrf   {V,_{r\f}}
\def \Vdd   {V_{22}}
\def \Vthth {V,_{\th\th}}
\def \Vdt   {V_{23}}
\def \Vtf   {V,_{\th\f}}
\def \Vtt   {V_{33}}
\def \Vff   {V,_{\f\f}}
\def \Wn   {W_{0}}
\def \Wu   {W,_{u}}
\def \Wj   {W_{1}}
\def \Wr   {W,_{r}}
\def \Wd   {W_{2}}
\def \Wth  {W,_{\th}}
\def \Wt   {W_{3}}
\def \Wf   {W,_{\f}}
\def \Wjj   {W_{11}}
\def \Wrr   {W,_{rr}}
\def \Wjd   {W_{12}}
\def \Wrt   {W,_{r\th}}
\def \Wjt   {W_{13}}
\def \Wrf   {W,_{r\f}}
\def \Wdd   {W_{22}}
\def \Wthth {W,_{\th\th}}
\def \Wdt   {W_{23}}
\def \Wtf   {W,_{\th\f}}
\def \Wtt   {W_{33}}
\def \Wff   {W,_{\f\f}}
\def \Un  {U_{0}}
\def \Uu  {U,_{u}}
\def \Uj  {U_{1}}
\def \Ur  {U,_{r}}
\def \Ud  {U_{2}}
\def \Uth {U,_{\th}}
\def \Ut  {U_{3}}
\def \Ufi  {U,_{\f}}
\def \Ujj   {U_{11}}
\def \Urr   {U,_{rr}}
\def \Udd   {U_{22}}
\def \Uthth {U,_{\th\th}}
\def \Udt   {U_{23}}
\def \Utf   {U,_{\th\f}}
\def \Utt   {U_{33}}
\def \Uff   {U,_{\f\f}}
\def \Ujd   {U_{12}}
\def \Urt   {U,_{r\th}}
\def \Ujt   {U_{13}}
\def \Urf   {U,_{r\f}}

\def \sn {\sin \th}
\def \cs {\cos \th}
\def \tg {\tan \th}
\def \ctg {\cot \th}
\def \csec {\csc \th}
\def \dcsec {\csc^2 \th}
\def \msn {\sin^{-1} \th}
\def \ct {\bar t}
\def \dsn {\sin^2 \th}
\def \mdsn {\sin^{-2} \th}
\def \tsn {\sin^3 \th}
\def \csn {\sin^4 \th}
\def \dcs {\cos^2 \th}
\def \tcs {\cos^3 \th}
\def \csin {\sqrt{1-(wu)^2}}
\def \dctg {\cot^2 \th}
\def \chd  {\cosh 2\d}
\def  \shd  {\sinh 2\d}
\def \mm {\mbox{\quad }}
\def \mv {\mbox{\qquad }}
\def \msip {\rightarrow}
\def \vsip {\longrightarrow}
\def \lkz  {\bigl(}
\def \pkz  {\bigr)}
\def \lvkz {\Bigl(}
\def \pvkz {\Bigr)}
\def \lvvkz {\biggl(}
\def \pvvkz {\biggr)}
\def \lhz  {\bigl[}
\def \phz  {\bigr]}
\def \lvhz {\Bigl[}
\def \pvhz {\Bigr]}
\def \lvvhz {\biggl[}
\def \pvvhz {\biggr]}
\def \lsz   {\bigl\{ }
\def \psz   {\bigr\} }
\def \pvsz {\Bigl\} }
\def \lvsz {\Bigr\{ }
\def \lvvsz {\Biggl\{}
\def \pvvsz {\Biggr\}}
\newcommand{\zl}[2]{{{\scriptstyle{\frac{#1}{#2}}}}}
\def \pul {{{\scriptstyle{\frac{1}{2}}}}}
\def \tripul {{{\scriptstyle{\frac{3}{2}}}}}
\def \ctvrt {{{\scriptstyle{\frac{1}{4}}}}}
\def \osmina {{{\scriptstyle{\frac{1}{8}}}}}
\def \sestina {{{\scriptstyle{\frac{1}{6}}}}}
\def \B  {\pul B}
\def \V  {\frac{V}{r}}
\def \AB {(A-\B)}
\def \AnpB {\An+\B}
\def \AnmB {\An-\B}
\def \ABV {\lvkz A\emdb-\frac{1}{2r}BV\pvkz}
\def \ee {(e,_\th +e\ctg)}
\def \cc {(c,_\th+2c\ctg)}
\def \dd {(d,_\th+2d\ctg)}
\def \ff {\frac{\tilde{f}}{r}}
\def \gg {\frac{\tilde{g}}{r\sn }}
\def \FUW  {(\Fnj-U\Fjd-W\Fjt\csec)}
\def \VUW {\left(\V\edb-r^2\edg U^2\chd
           -r^2\emdg W^2\chd-2r^2 UW\shd\right)}
\def \Uff {\lvkz BU+\ffg\pvkz}
\def \ffg {\frac{\tilde{f}\emg\sqrt{ch}}{r}}
\def \ggg {\frac{\tilde{g}\eg}{r\sn}}
\def \gggb {\frac{\tilde{g}\eg}{r\sqrt{ch}}}
\def \ffc  {(\fn-\fmj c)}
\def \Bffc {\lvkz-B\cc+\fj-\fn c+\pul\fmj c^2\pvkz}
\def \edb  {{\rm e}^{2\b }}
\def \emdb {{\rm e}^{-2\b }}
\def \eb   {{\rm e}^{\b }}
\def \emb  {{\rm e}^{-\b }}
\def \edg  {{\rm e}^{2\g }}
\def \emdg {{\rm e}^{-2\g }}
\def \eg   {{\rm e}^{\g }}
\def \emg  {{\rm e}^{-\g }}
\def \edbmg {{\rm e}^{2(\b -\g)}}
\def \edgmb {{\rm e}^{2(\g -\b)}}
\def \edbg  {{\rm e}^{2(\b +\g)}}
\def \edgb  {{\rm e}^{2(\g +\b)}}
\def \emdbg  {{\rm e}^{-2(\b +\g)}}
\def \co    {(2ch^2-1)}
\def \Kmn {K_{\m\n}}
\def \Knn {K_{00}}
\def \Knj {K_{01}}
\def \Knd {K_{02}}
\def \Knt {K_{03}}
\def \Kjj {K_{11}}
\def \Kjd {K_{12}}
\def \Kjt {K_{13}}
\def \Kdd {K_{22}}
\def \Kdt {K_{23}}
\def \Ktt {K_{33}}
\def \GGmns {G_{\m\n\s}}
\def \GGmnl {G_{\m\n\l}}
\def \GGjdt {G_{123}}
\def \GGnjd {G_{012}}
\def \GGnjt {G_{013}}
\def \GGndt {G_{023}}
\def \Jm {J^{\m}}
\def \Jn {J^0}
\def \Jj {J^1}
\def \Jd {J^2}
\def \Jt {J^3}
\def \Er {E^\r}
\def \Ef {E^\f}
\def \Ez {E^z }
\def \Br {B^\r}
\def \Bf {B^\f}
\def \Bz {B^z }
\def \Ert {E^{(\r)}}
\def \Eft {E^{(\f)}}
\def \Ezt {E^{(z )}}
\def \Brt {B^{(\r)}}
\def \Bft {B^{(\f)}}
\def \Bzt {B^{(z )}}
\def \Rma {R_{\m \a }}
\def \Rmb {R_{\m \b }}
\def \Rae {R_{\a \ep }}
\def \Rab {R_{\a \b }}
\def \Rmd {R_{\m \d }}
\def \Rmn {R_{\m \n }}
\def \Rnn {R_{00}}
\def \Rnj {R_{01}}
\def \Rnd {R_{02}}
\def \Rnt {R_{03}}
\def \Rjn {R_{10}}
\def \Rjj {R_{11}}
\def \Rjd {R_{12}}
\def \Rjt {R_{13}}
\def \Rdn {R_{20}}
\def \Rdj {R_{21}}
\def \Rdd {R_{22}}
\def \Rdt {R_{23}}
\def \Rtn {R_{30}}
\def \Rtj {R_{31}}
\def \Rtd {R_{32}}
\def \Rtt {R_{33}}
\def \Ruab {R^{\a \b }}
\def \Ruae {R^{\a \ep }}
\def \Rumn {R^{\m \n }}
\def \Runn {R^{00}}
\def \Runj {R^{01}}
\def \Rund {R^{02}}
\def \Runt {R^{03}}
\def \Rujn {R^{10}}
\def \Rujj {R^{11}}
\def \Rujd {R^{12}}
\def \Rujt {R^{13}}
\def \Rudn {R^{20}}
\def \Rudj {R^{21}}
\def \Rudd {R^{22}}
\def \Rudt {R^{23}}
\def \Rutn {R^{30}}
\def \Rutj {R^{31}}
\def \Rutd {R^{32}}
\def \Rutt {R^{33}}
\def \Gdlae {\G^{\d }_{\a \ep }}
\def \Gdlab {\G^{\d }_{\a \b }}
\def \Gnyae {\G^{\n }_{\a \ep }}
\def \Gnyab {\G^{\n }_{\a \b }}
\def \Gnae {\G^{0}_{\a \ep }}
\def \Gnab {\G^{0}_{\a \b }}
\def \Gjae {\G^{1}_{\a \ep }}
\def \Gjab {\G^{1}_{\a \b }}
\def \Gdae {\G^{2}_{\a \ep }}
\def \Gdab {\G^{2}_{\a \b }}
\def \Gnmn {\G^{0}_{\m \n }}
\def \Gjmn {\G^{1}_{\m \n }}
\def \Gdmn {\G^{2}_{\m \n }}
\def \Gnjj {\G^{0}_{11}}
\def \Gdjj {\G^{2}_{11}}
\def \BE {\begin{equation}}
\def \EE {\end{equation}}
\def \BDM {\begin{displaymath}}
\def \EDM {\end{displaymath}}
\def \BEAH {\begin{eqnarray*}}
\def \EEAH {\end{eqnarray*}}
\def \BEA {\begin{eqnarray}}
\def \EEA {\end{eqnarray}}
\def \BM {\begin{math}}
\def \EM {\end{math}}
\author{V. Pravda  \footnote{E-mail: {\tt pravda@math.cas.cz} }, A. Pravdov\' a  \footnote{E-mail: {\tt pravdova@math.cas.cz} }}
\address{Mathematical Institute,\protect\\ 
Academy of Sciences, \protect\\ 
\v Zitn\' a 25,
115 67 Prague 1, Czech Republic }
\date{\today}
\maketitle
\begin{abstract}
Boost-rotation symmetric spacetimes are the~only locally asymptotically flat
axially symmetric electrovacuum spacetimes with a further symmetry
that are radiative. They are realized by uniformly accelerated particles of 
various kinds or black holes. Their general properties are summarized. Several examples
of boost-rotation symmetric solutions of the~Maxwell and Einstein equations
are studied: uniformly accelerated electric and magnetic multipoles,
the~Bonnor-Swaminarayan solutions, the~C-metric and the~spinning C-metric.
\end{abstract}
\pacs{PACS: 04.20.Jb, 04.30.-w, 04.20.Ha }
\tableofcontents
\section{ Introduction}

There exists only one class of exact radiative 
solutions of the~full nonlinear Einstein equations which 
are known in an analytical form, realized by
moving objects and,  in addition, are asymptotically flat 
(in some cases  null infinity $\J$ is even  global \cite{AstSchm}).
This is the~class of boost-rotation symmetric spacetimes
describing uniformly accelerated particles or black holes 
symmetrically located along the~symmetry axis
(see reviews  \cite{ujibiLes,JibiEhlers,JibiWeimar}).
They have two Killing vectors,
the~axial Killing vector with circular group orbits
and the~boost Killing vector which asymptotically
goes over
to the~generator of the~Lorentz transformation along the~symmetry axis 
in the~Minkowski spacetime.
The~significance of these solutions follows from the~theorem
 which states that among locally asymptotically
flat spacetimes with the~axial and an additional symmetry, 
boost-rotation symmetric 
spacetimes are the~only spacetimes which are radiative. It was  proved in \cite{jibisch} for vacuum
spacetimes with
hypersurface orthogonal Killing vectors and generalized in \cite{ajajibi} for electrovacuum
spacetimes with Killing vectors which are not hypersurface orthogonal. 
These spacetimes
have been found also useful in numerical relativity 
as test beds for numerical codes (approaches 
based on null \cite{wini,gomez} and spacelike \cite{alcu} initial hypersurfaces).

Several boost-rotation symmetric solutions are known.
 The~first boost-rotation symmetric solution, that has been found,
 is the~C-metric  \cite{Levi,Weyl}, describing accelerated
black holes symmetrically located on the~axis of symmetry  with
a nodal (conical) singularity (``cosmic string'') which causes the~acceleration.
Another example is a set of Bonnor-Swaminarayan solutions \cite{BSZ}
representing a finite number of uniformly accelerated monopole ``Curzon-Chazy''  particles.
The~acceleration can be caused by nodal singularities or
by  mutual gravitational interaction.
However, in the~case with no  nodal singularities, negative masses 
occur. The~simplest case without  conical singularity is
represented by two pairs of particles, in each of which 
there is one particle with positive and  one with
negative mass (as an analogous system
in the~Newton mechanics, such a system is self-accelerated). 
A limiting procedure (similar to that in electromagnetism by 
which one obtains a dipole from two monopoles) leads to a special
solution \cite{jibiHS} realized by two independent,
self-accelerating particles. Such a solution
is asymptotically flat, admitting global null infinity
${\cal J}$. 
Boost-rotation symmetric solutions with a cosmic
string extending along the~whole axis of symmetry are 
locally asymptotically flat 
in the~sense that they admit only local ${\cal J}$. 
Boost-rotation symmetric solutions 
describing uniformly accelerated particles with a general 
multipole structure  were also found in \cite{jibiHS}. 
There exist   generalizations of the~vacuum boost-rotation
symmetric spacetimes containing 
an electromagnetic field 
-- the~charged C-metric \cite{uKinnWalker}, or rotating sources -- 
the~spinning C-metric \cite{PlebDem,bivoj}. The~Killing vectors in the~case
of a spinning source are not hypersurface orthogonal. 
Also the~generalized \mbox{C-me\-tric} \cite{Ernst78}, generalized charged C-metric \cite{Ernst76},
and generalized Bonnor-Swaminarayan solution \cite{jibiHS} are known that are not asymptotically flat
and in which an external field is present to cause the~acceleration. 

Section \ref{sec2} is devoted to the~general theory of boost-rotation symmetric
spacetimes. It is divided into two parts. In the~first part  \ref{sub21Th}
we recall the~Bondi method we further use to present the~theorem 
about uniqueness of  boost-rotation symmetric spacetimes among all
locally asymptotically flat electrovacuum axially symmetric spacetimes with an additional symmetry.
In the~second part \ref{secgen}, based on \cite{BicSchPRD},  the~definition and general features of boost-rotation symmetric spacetimes are summarized. 

Sec.~\ref{sec3} discusses several examples of boost-rotation symmetric solutions:
uniformly accelerated electric and magnetic multipoles, the~Bonnor-Swaminarayan solutions,
the~C-metric and the~spinning C-metric.

We use the~signature $-2$ everywhere except for the~last two parts
concerning the~standard and spinning C-metrics where the~signature $+2$
is used following \cite{PlebDem,bivoj}. 


\section{Theoretical background}
\label{sec2}

\subsection{Asymptotically flat axisymmetric spacetimes and radiation}
\label{sub21Th}

\subsubsection{The~Bondi method}
\label{secbondi}

If one is interested in  gravitational radiation 
from a general bounded matter source, i.e., in  the~behaviour
of  gravitational field far from the~source, one has to turn
to approximation methods, 
typically one expands the~metric
in negative powers of a suitably chosen ``radial coordinate'' $r$.
This was done for axially symmetric non-rotating sources by Bondi {\it et al.} \cite{bondi}
  (generalized without this assumption by Sachs \cite{sachs},  for charged sources
by van der Burg \cite{burg} -- mistakes were corrected in \cite{ajajibi},
for spacetimes with null dust  by von der G\"{o}nna and Kramer  \cite{gonna} and for spacetimes
with polyhomogeneous $\J$ by Chru\' sciel {\it et al.} \cite{Chrusciel}).  
They introduced suitably chosen coordinates
$\{ u,\ r,\ \theta,\ \phi \}$ in such a way that, roughly speaking, 
null coordinate
 $u$, spherical angles $\theta$ and $\phi$ are constant along outgoing radial null 
geodesics -- light rays -- meanwhile  $r$, the~{\it luminosity distance},
varies and satisfies the~condition $g_{\theta \theta} g_{\phi \phi} = r^4 \sin^2 \theta$
so that the~area of the~surface $u=$ const, $r=$ const, $\phi\in\langle 0,2\p)$,
and $\theta\in\langle 0,\p\rangle$ is $4\pi r^2$. 

The~Bondi  method  meant a breakthrough in gravitational radiation theory. 
It consists in prescribing initial data not on a spacelike Cauchy
hypersurface and solving the~Cauchy problem (which is well posed
for the~Einstein equations \cite{Wald}) 
but in prescribing initial data on a characteristic hypersurface
of the~quasilinear hyperbolic Einstein equations which is {\it null},
i.e.,  on a hypersurface $u=$ const. 
Then  discontinuities
can arise and the~Cauchy problem is ambiguous.
One has  to prescribe initial data  not only on the~initial null
hypersurface but one also has to know two functions, 
$c,_u$ and $d,_u$
(in the~vacuum case), for all $u$ (or $u$ in the~given interval).
Only then the~evolution of the~gravitational field is known, i.e., determined for all $u$ 
(or in the~given interval). This is why we call these two functions
the~{\it news functions} of the~system 
because they carry the~whole information
about changes in the~system. If and only if one of them does not
vanish the~total mass of the~system as measured
at null infinity, the~{\it Bondi mass}, decreases and gravitational
waves are radiated.
If the~$u$-derivative of at least one of the~news functions
does not vanish the~Weyl tensor has radiative tetrad 
components (proportional to $r^{-1}$)
\BE
C_{\a\b\g\d}m^\a t^\b m^\g t^\d
              =[(c+{\rm i}d)_{,uu}]\frac{1}{r}+{\cal O}(r^{-2})\ 
         \label{Weyl}
\EE
in the~standard null tetrad $\{k^\a ,\ m^\a ,\ t^\a ,\ {\bar t}^\a\} $ introduced in \cite{sachs,ajajibi}.
There are exceptional cases 
corresponding, e.g., to an infinite string (thus the~axis of axial symmetry is not regular)
with a non-zero news function and with  a non-radiative Weyl tensor  \cite{ABS1}.

Hereafter  we are interested in asymptotically flat electrovacuum spacetimes
that are in addition axially symmetric but a source can rotate
 and we thus follow the~work of
van der Burg \cite{burg} with the~additional  assumption that all metric functions
do not depend on $\phi$.
The~metric has, in the~Bondi-Sachs coordinates \mbox{ \{ $u$,~$r$,~$\th$,~$\f$ \} $\rov$ \{
             $x^0$,~$x^1$,~$x^2$,~$x^3$\}}, the~form
\BEA
{\rm d}s^2&=&\VUW {\rm d}u^2\nn\\
    & &\ + 2\edb {\rm d}u {\rm d}r
              +2r^2\left(\edg U\chd +W\shd\right)  {\rm d}u{\rm d}\theta\nn\\
    & &    
        \ +2r^2\left(\emdg W\chd +U\shd\right)\sn\ {\rm d}u{\rm d}\phi\nn\\
    & &\ - r^2\left[ \chd\left(\edg {\rm d}\theta^2 +\emdg \dsn\ {\rm d}\phi^2 \right)
                        +2\shd\sn\ {\rm d}\theta {\rm d}\phi\right] \ ,\label{ds}
\EEA
where all the~six metric functions $U$, $V$, $W$, $\b$, $\g$, $\d$ and
the~electromagnetic field  $\Fmn$ do not depend on $\f$ because of the~axial
symmetry. If one requires the~axis of the~axial symmetry ($\th=0$, $\p$) to be regular,
then functions 
\BE
V, \mm  \b , \mm W, \mm \g/ \dsn , \mm U / \sn , \mm \d / \sn\label{regul}
\EE
have to be regular for $\sn \msip 0$.

First assuming the~functions $\g$,  $\d$,  $F_{12}$ and  $F_{13}$
to have asymptotic expansions, the~spacetime to be asymptotically flat,
and only the~outgoing radiation to be asymptotically present
then the~expansions of these four functions 
at large $r$ on an initial hypersurface $u=$ const are of the~form
\BEA
\g  &=&\frac{c}{r}+( C-{{\scriptstyle{\frac{1}{6}}}}c^3
             -\tripul cd^2)
           \frac{1}{r^3}+\frac{D}{r^4}
         +{\cal O}(r^{-5})\ ,\nn\\
\d  &=&\frac{d}{r}+
           (H-{{\scriptstyle{\frac{1}{6}}}}d^3
             +\pul c^2 d)\frac{1}{r^3}
                  +\frac{K}{r^4}+{\cal O}(r^{-5})\ ,\label{HRN}\\
\Fjd&=&\frac{e}{r^2}+(2E+ec+fd) \frac{1}{r^3}
             +{\cal O}(r^{-4})\ ,\nn\\  
\Fjt&=&\lvvhz\frac{f}{r^2}+(2F+ed-fc) \frac{1}{r^3}
              +{\cal O}(r^{-4})\pvvhz\sn\ .\nn
\EEA
Prescribing then initial data on this null hypersurface, i.e., $c$, $d$, 
$C$, ... and $M$, $N$,  $P$,  $\ep$,  $\m$  as functions of  $\th$, 
all the~other metric functions are determined on the~initial hypersurface by the~field equations:
\BEA
\b  &= & -\ctvrt (c^2+d^2)\frac{1}{r^2}+{\cal O}(r^{-4})\ ,
               \nn \\  
U   &= &-\cc\frac{1}{r^2}
        +\lvhz 2N+3(cc,_\th+dd,_\th)+4(c^2+d^2)\ctg\pvhz
           \frac{1}{r^3}\nn\\  
    & &+\pul\lvhz 3(C,_\th+2C\ctg)-6(cN+dP)
            -4(2c^2c,_\th+cdd,_\th+c,_\th d^2)
           -8c(c^2+d^2)\ctg
          +2(\ep e-f\m)\pvhz\frac{1}{r^4}+{\cal O}(r^{-5})\ ,\nn\\
W   &= &-\dd\frac{1}{r^2}
        +\lvhz 2P+2(c,_\th d-cd,_\th)\pvhz\frac{1}{r^3}\nn\\
    & &+\pul\lvhz 3(H,_\th+2H\ctg)+6(cP-dN)
                 -4(2d^2d,_\th+cdc,_\th+c^2d,_\th)
            -8d(c^2+d^2)\ctg+2(\m e+\ep f)\pvhz\frac{1}{r^4}
                          +{\cal O}(r^{-5})\ ,\nn\\
V   &= &r-2M
        -\lvhz N,_\th+N\ctg-\pul (c^2+d^2) 
         -\cc^2-\dd^2 -(\ep^2+\m^2)\pvhz\frac{1}{r}\nn\\
    & &-\pul \lvhz 
       C,_{\th\th}+3C,_\th\ctg-2C+6N\cc
                                 +6P\dd 
        +4(2cc,_\th^2+3c,_\th dd,_\th-cd,_\th^2)  \nn\\  & &
           +8(2c,_\th d^2+3c^2c,_\th+cdd,_\th)\ctg
          +16c(c^2+d^2)\dctg
             +2\ep\ee-2\m (f,_\th+f\ctg)\pvhz\frac{1}{r^2}
                +{\cal O}(r^{-3})\ ,\nn\\  
\Fnj&=&-\frac{\ep}{r^2}+( e,_\th+e\ctg)\frac{1}{r^3}
          +{\cal O}(r^{-4})\ ,\label{dlouhyrozvojmetriky}\\   
\Fdt&=&\lvhz -\m-( f,_\th+f\ctg)\frac{1}{r}+{\cal O}(r^{-2})\pvhz\sn\ ,\nn\\
\Fnd&=&X+(\ep,_\th-e,_u)\frac{1}{r}
        -\lvsz [E+\pul (ec+fd)],_u
              +\pul\ee,_\th\pvsz\frac{1}{r^2}
                  +{\cal O}(r^{-3})\ ,\nn\\   
\Fnt&=&\lvvsz Y-\frac{f,_u}{r}
        -\lvhz [F+\pul(ed-fc)],_u\pvhz\frac{1}{r^2}
                   +{\cal O}(r^{-3})\pvvsz\sn\ .\nn
\EEA
Provided  four news functions, two gravitational, $c,_u$, $d,_u$, and two electromagnetic, $X$, $Y$, 
are prescribed for all $u$, the~time evolution of the~system is fully determined using the~Einstein-Maxwell
equations (see \cite{burg} for details). Let us just mention a  relation for 
the~{\it mass aspect} $M$
which is equal in stationary case to the~total mass of the~system:
\BE
M,_u  =-(c,_u^2+d,_u^2)-(X^2+Y^2)+\pul(c,_{\th\th}+3c,_\th\ctg-2c),_u\ .\label{Mu}
\EE

Then the~total mass at future null infinity $\J^+$ at a given ``retarded time'' $u$, 
the~Bondi mass $m$, defined 
as a mean value of the~mass aspect of the~system, $M$, 
over the~sphere,\footnote{If the~spacetime is stationary,  the~Bondi mass $m$
is the~same as the~mass aspect $M$ and is equal to the~total mass of the~system as measured at spatial
infinity -- the~ADM mass.}
\BE
m(u)=\pul\int_{0}^{\p} M(u,\th)\sn {\rm d}\th\ ,\label{hmota}
\EE
is decreasing if at least one of the~news functions does not vanish:
\BE
m,_u =-\pul\int\limits_{0}^{\p} (c,_u^2+d,_u^2+X^2+Y^2)\sn {\rm d}\th
               \leq 0\ .  \label{klhmota}
\EE
The~mass decrease  is caused by  
gravitational and electromagnetic waves radiated out  from  the~system.

The~rate of loss of the~electromagnetic energy
radiated out from the~system is given by
\BE
\frac{\der}{\der u}\int\limits_{0}^{\p} \pul (X^2+Y^2)\sn {\rm d}\th\ ,
\label{tokelmgen}
\EE
which indeed implies the~loss of mass as seen from Eq.~(\ref{klhmota}).

There are several conserved quantities described in  \cite{burg,burg9}.
Two of them are the~electric and magnetic charges of the~source
\BE
\frac{\der}{\der u}\int\limits_{0}^{\p}\pul\ep\sn {\rm d}\th =0\ ,\mm\mm
\frac{\der}{\der u}\int\limits_{0}^{\p}\pul\m\sn {\rm d}\th =0\ .
\label{zachnab}
\EE

\subsubsection{Symmetries of asymptotically flat axisymmetric electrovacuum spacetimes 
and radiation}

Since finding an exact  radiative solution describing a general charged bounded
source is a very difficult task, may be unsolvable, it is of interest to know
which isometries are compatible with asymptotic flatness and admit
radiation. 
This question was solved in \cite{jibisch}  for locally asymptotically flat
axially symmetric vacuum spacetimes with non-rotating sources, i.e.,
with hypersurface orthogonal Killing vectors, and generalized in \cite{ajajibi}
for electrovacuum spacetimes with, in general, rotating sources,
i.e., with Killing vectors that are not hypersurface orthogonal.
The~assumption of axial symmetry with the~corresponding
Killing vector field denoted by $\x=\der /\der\f$ simplifies lengthy calculations.

Suppose that such an axially symmetric electrovacuum
spacetime admits only a ``piece'' of future null infinity $\J^+$ in the~sense
of \cite{AstSchm}. Then one can introduce the~Bondi-Sachs coordinate
system \mbox{ \{ $u$,~$r$,~$\th$,~$\f$ \} $\rov$ \{
             $x^0$,~$x^1$,~$x^2$,~$x^3$\}}
in which the~metric has the~form (\ref{ds}), where the~metric functions
and the~electromagnetic field $\Fmn$ are functions of $u$, $r$, $\th$ and 
have expansions (\ref{HRN}), (\ref{dlouhyrozvojmetriky}). 
Thanks to the~existence of only local $\J^+$, we assume 
Eqs.  (\ref{HRN})--(\ref{Mu}) 
to be satisfied  
for all $\f\in\langle 0,2\p) $, however, 
not necessarily  on the~whole sphere, i.e., not for all $\th\in\langle 0,\p\rangle$
but only in some open interval of $\th$ and thus the~Bondi mass (\ref{hmota})
cannot be introduced. For example, 
the~``axis of symmetry'' ($\th=0$, $\p$) may contain
nodal singularities and thus need not be regular and the~regularity
conditions on the~axis (\ref{regul}) need not be satisfied for any $u$.
If the~axis is singular then at least two generators of $\J^+$
would be missing so that $\J^+$ would not be topologically $S^2\times{\cal R}$.

Let us now assume that another Killing vector field $\e$ exists
which forms together with $\x =\der / \der\f$ a two-parameter
group. In Ref. \cite{jibisch}
it is  proved (see Lemma in Sec. 2) that in the~case of $\x$ with
circles as integral curves, $\x$ and $\e$ determine an Abelian
Lie algebra so that we can assume $[\e,\x]=0$. Hence,
the~components  $\e^\a$ are independent of $\f$.

Then by solving the~Killing equations $ \L_\e \gab=0 $ 
(explicitly given in \cite{ajajibi,ajadis}), 
the~Killing vector asymptotically turns out to be 
\BE
\e^\a=[-ku\cs+\a(\th),\
          kr\cs+{\cal O}(r^{0}),\ -k\sn+{\cal O}(r^{-1}),\ {\cal O}(r^{-1})]\ , 
\label{Bbotr}
\EE
where $k$ is a constant and  $\a$ an arbitrary function of $\th$.
This result for a locally  asymptotically flat axially symmetric electrovacuum
spacetime with Killing vectors which need not be hypersurface
orthogonal proved in \cite{ajajibi} is the~same as that one for asymptotically
flat axially symmetric vacuum spacetimes with hypersurface
orthogonal Killing vectors obtained in \cite{jibisch}.

Let us analyze  possibilities $k=0$ and $k\not= 0$ separately.\\[1mm]
1) {\it Case $k=0$}\\[1mm]
This alternative was considered in \cite{ajajibi} and developed in detail in \cite{ajajibi2}.
When \mbox{$k=0$}, the~Killing vector field $\e$ (\ref{Bbotr}) generates
supertranslations.
 However, solving Killing equations in higher orders
of $r^{-1}$ and assuming that the~electromagnetic field described
by $\Fmn$ (\ref{HRN}), (\ref{dlouhyrozvojmetriky}) has the~same symmetry, i.e., $\L_\e \Fab=0$,
one can show that $\e$ in fact generates translations
\BEA
\e^\a&=&\lvhz B,\ 
         Bc,_u+B,_\th\ctg-\frac{uc,_u^2B}{r}           +{\cal O}(r^{-2}),
       \ -B,_\th \frac{1}{r}+B,_\th\frac{c}{r^2}+{\cal O}(r^{-3}),
\ {\cal O}(r^{-4})\pvhz\ ,
            \label{etasupertrans}
\EEA
where
\BE
B(\th)=\frac{\sn}{2{\rm C}}\lvvhz (b_0+a_0)\lvkz\frac{\sn}{\cs+1}\pvkz^{\rm C}
 +(b_0-a_0)\lvkz\frac{\sn}{\cs+1}\pvkz^{-\rm C}\pvvhz \ ,\label{strB}
\EE
with constants $a_0$, $b_0$ and ${\rm C}\in ( 0,1\rangle$ in a spacetime
with, in general, an infinite thin cosmic string along the~symmetry axis. 
The~cosmic string is described by a deficit angle $2\p (1-{\rm C})$.
The~only non-vanishing news function -- due to the~presence of the~string -- is
\BE
c,_u=\frac{{\rm C}^2-1}{2\dsn}\  ,
\EE
which depends only on  $\th$ and  thus the~Weyl tensor (\ref{Weyl}) does not
have a radiative character. The~leading metric functions and electromagnetic
field functions have the~form 
\BEA
c(u,\ \th) &=& \frac{u}{2B}(B,_{\th\th}-B,_\th\ctg)
  = u\frac{{\rm C}^2-1}{2\dsn}\ ,\mm d(u,\ \th)=0\ ,\nn\\
M(u,\ \th) &=& -uc,_u^2-a_1 B^{-3}\ ,\nn\\
N(u,\ \th) &=& -\pul B^{-4}(2a_1B,_\th u+f_0\sn)\ ,\mm 
       P(u,\ \th)=-\pul g_0\sn B^{-4}\ ,\label{fcetransl}\\
X(u,\ \th) &=& Y(u,\ \th)=0\ ,\mm
\ep(u,\ \th)=\ep_0B^{-2}\ ,\mm \m(u,\ \th)=\m_0 B^{-2}\ ,\nn\\
e(u,\ \th) &=& -\ep_0 B,_\th B^{-3}u+ e_1(\th)\ ,\mm
f(u,\ \th)=\m_0 B,_\th B^{-3}u+f_1 (\th)\ ,\nn
\EEA
where $a_1$, $f_0$, $g_0$, $\ep_0$, and $\m_0$ are constants and $e_1$, $f_1$ are arbitrary
functions of $\th$.
The~reader may find more metric and field functions in  
\cite{ajadis} and \cite{ajajibi2}.

In paper \cite{ajajibi2} we proved
the~following statements in Theorems 1, 2.
The~asymptotically translational Killing vector $\e$, given by (\ref{etasupertrans}),
has asymptotically the~norm 
\BEA
\| \e\|^2&=&(b_0^2-a_0^2)+\frac{2a_1}{B}\frac{1}{r}
         +[ u2a_1B^{-3}(B^2c,_u-B,_\th^2+BB,_\th\ctg)
        \nn\\  &&\mm
           +f_0\sn\ B^{-3}(-B,_\th+B\ctg)+(\ep_0^2+\m_0^2)B^{-2}]
             \frac{1}{r^2}+{\cal O}(r^{-3})\ ,
\label{norma}
\EEA
that may be spacelike, timelike or null. 
It is spacelike   for $a_0^2>b_0^2$.
If one of the~constants  $a_1$,  $f_0$ or $g_0$ is non-vanishing then
 $\J^+$ is singular at $\th=\th_0\not= 0$, $\p$, given by the~relation
\BE
\cos\th_0=\frac{(a_0+b_0)^{2/{\rm C}}-(a_0^2-b_0^2)^{1/{\rm C}}}
               {(a_0+b_0)^{2/{\rm C}}+(a_0^2-b_0^2)^{1/{\rm C}}}\ ,
\label{vlnydivergence}
\EE
 in addition to the~singularity due to the~presence 
of the~cosmic string at $\th=0$, $\p$. 
This case corresponds
to cylindrical waves (in particular, $\th_0=\p /2$ for $b_0=0$).
The~Killing vector $\e$ is  null if $a_0=\pm b_0$.
Then $\J^+$ is singular at $\th=0$ or $\p$ in addition to the~string singularity
if again $a_1$ or $f_0$ or $g_0$ is non-vanishing. It corresponds
to a wave propagating along the~symmetry axis. 
The~Killing vector $\e$ is timelike for $a_0^2<b_0^2$
 and the~only singularities of $\J^+$ at $\th=0$, $\p$  
are due to the~presence of the~string.

If there is no string, null infinity $\J^+$ may be regular  
even with a non-vanishing Bondi mass $m$
for an asymptotically timelike translational Killing vector.
However, even  if the~translational Killing vector is spacelike or null, 
$\J^+$ may be regular if the~spacetime is flat in its neighbourhood 
(i.e., constants $a_1$, $f_0$, $g_0$ and thus also the~mass aspect $M$ and
functions $N$, $P$, ... vanish).

For $\rm C=1$, without the~string, we find the~Killing vector field
$\e$ (\ref{etasupertrans}) to have asymptotically the~form 
\BE
\e^\a=\lvhz -a_0\cs+b_0,\ a_0\cs+{\cal O}(r^{-2}),\ -\frac{a_0\sn}{r}+{\cal O}(r^{-4}),\
{\cal O}(r^{-4})\pvhz
\EE
and the~mass aspect  
is then given by
\BE
M=-\frac{a_1}{(-a_0\cs+b_0)^3}\ .
\label{hmotabezstruny}
\EE
In the~case of the~timelike
Killing vector ($b_0^2>a_0^2$) assuming $b_0>0$ and $a_1<0$
we obtain the~total Bondi mass to be
\BE
m=-\frac{a_1b_0}{(b_0^2-a_0^2)^2}>0\ . \label{mhotabezstruny}
\EE
The~factor \mbox{$-a_0\cs+b_0$} appearing in
Eq.~(\ref{hmotabezstruny}) 
corresponds, using the~terminology of \cite{bondi},  
to the~``Doppler shift of the~mass aspect'' and the~other
Bondi functions. It occurs when the~system 
is boosted with respect
to the~Bondi frame with the~boost parameter -- $\n$ ($b_0=\cosh \n$, $a_0=\sinh \n$) so that its velocity
is $v=-\tanh \n=a_0/b_0$. 
Putting $a_1 = - {\bar m}$, we get 
the~mass aspect in Eq.~(\ref{hmotabezstruny}) in the~form
$M = {\bar m}/ (\cosh\n - \sinh\n \cs)^3$,
which exactly corresponds to the~formula 
(see Eq.~(72) in \cite{bondi}) for the~Schwarzschild 
mass ${\bar m}$ moving along the~axis of symmetry with
constant velocity $-\tanh\n$.

In \cite{ajajibi2} two exact solutions are studied  with
axial and translational Killing vectors and with an infinite
cosmic string along the~symmetry axis -- the~Schwarzschild solution
with a string, where the~Killing vector  generates translations along \mbox{the~$t$-axis}
(\mbox{$a_1\not=0$}, \mbox{$f_0=g_0=0$})
and Einstein-Rosen cylindrical waves with a string having a spacelike
translational Killing vector corresponding to translations along the~$z$-axis
($a_1=0$, $f_0\not= 0$, $g_0=0$).\\[1mm]
2) {\it Case $k\not= 0$}\\[1mm]
Assuming $k\not= 0$ in (\ref{Bbotr}), it is easy to find
a Bondi-Sachs coordinate system with $\a=0$ by performing a supertranslation.
Hence we put $\a=0$ in (\ref{Bbotr}) and without loss of generality
we choose $k=1$. Then the~asymptotic form of the~Killing vector field $\e$
is
\BE
\e^\a=[-u\cs,\ r\cs+{\cal O}(r^0),\ -\sn+{\cal O}(r^{-1}),\ {\cal O}(r^{-1}) ]\ .
\EE
It is the~{\it boost} Killing vector that generates the~Lorentz transformation
along the~axis of axial symmetry.

Solving the~Killing equations in higher orders of $r^{-1}$ and assuming that the~electromagnetic
field has the~same symmetry, i.e., $\L_\e \Fmn=0$ is 
satisfied, we obtain the~forms of the~metric and field functions to be:
\BEA
c(u,\th)&=&-\frac{\K(w)}{uw}\mm\msip\  c,_u=\frac{\K(w),_w}{u^2}\ ,\label{boost-c}\\
d(u,\th)&=&-\frac{\L(w)}{uw}\mm\msip\  d,_u=\frac{\L(w),_w}{u^2}\ ,\label{boost-d}\\
M(u,\th)&=&\frac{1}{2\sn}(w^2\K,_w),_w+\frac{\l(w)}{u^3w^3}\ ,\label{boost-M}\\
X(u,\th)&=&\frac{\E(w)}{u^2}\ ,\label{boost-X}\\
Y(u,\th)&=&\frac{\BB(w)}{u\sn}=\frac{\BB(w)}{wu^2}=\frac{\tilde{\BB}(w)}{u^2}
    \ ,\label{boost-Y}\\
\ep(u,\th)&=&\frac{\E(w)}{  u}\ctg\ ,\label{boost-ep}\\
\m(u,\th) &=&\frac{\BB(w)}{\sn}\ctg = \frac{\tilde{\BB}(w)}{  u}\ctg\ ,\label{boost-mu}\\
e(u,\th)&=&\frac{\E(w)}{2}+\frac{\GG (w)}{u^2}\ ,  \label{boost-e}\\
f(u,\th)&=&-\frac{\BB (w)u}{2\sn}+\frac{\D (w)}{u\sn}
  = -\frac{\tilde{\BB}(w)}{2}+\frac{\tilde{\D}(w)}{u^2}\ , \label{boost-f}
\EEA
where $w=\sn /u$, functions $\K(w)$, $\L(w)$,  $\E(w)$ and $\BB(w)$ ($\tilde{\BB}(w)=\BB(w)/w$)
are arbitrary functions determining news functions, 
and  $\l(w)$ has to satisfy the~equation
\BE
\l,_w=w^2(\K,_w^2+\L,_w^2+\E^2+\tilde{\BB}^2)
-\frac{1}{2w}(w^3\K,_{ww}),_w\ ,\label{boostlambda}
\EE
while $\GG(w)$ and $\D(w)$ ($\tilde{\D}(w)=\D (w)/w$)
are solutions of the~equations 
\BEA
2w^2(\GG,_w w+2\GG)+(\E,_w w-\E)+2w\E\K+2\BB\L\hspace*{6mm}  &=& 0\ ,\\
2w^2(\D,_w w+\D)-(\BB,_w w-2\BB)-2w\BB\K+2w^2\E\L &=& 0\ .
\EEA
Solving Eq.~(\ref{boostlambda}) for $\l$  with prescribed news functions
(i.e., for given $\K(w),_w$, $\L(w),_w$,  $\E(w)$ and $\tilde{\BB}(w)$)
 we find the~mass aspect $M(u,\ \th)$
(\ref{boost-M}) and thus the~total Bondi mass at $\J^+$ is then given by
\BE
m(u)=\pul\int\limits_{0}^{\p} M(u,\th)\sn \ {\rm d}\th\
     =\ctvrt\int\limits_0^\p (w^2\K,_w),_w{\rm d}\th
     +\pul\int\limits_0^\p \frac{\l}{u^2w^2}\ {\rm d}\th\ .\label{boostm}
\EE
The~expansion of the~boost Killing vector is as follows
\BEA
\e^\a&=&
      \lvhz -u\cs\ ,\ r\cs+u\cs+\cs\ \lvkz\K,_w +\frac{\K}{w}\pvkz
                      \ \frac{1}{r}
              +{\cal O}(r^{-2})\ ,\ \nn\\
     &&\ 
             -\sn- \frac{u\sn}{r} + \frac{uc\sn}{r^2}+{\cal O}(r^{-3})\ ,
               \  \frac{ud}{r^2}+{\cal O}(r^{-3})\pvhz\ .\label{etaboost}
\EEA

These results were generalized for spacetimes with in general polyhomogeneous $\J$
using Newman-Penrose formalism in \cite{Kroon} .

Since in boost-rotation symmetric spacetimes the~news functions
are in general non-vanishing functions of $u$ and $w$, these spacetimes are radiative
and so we may conclude with the~theorem proved in \cite{ajajibi}:\\[2mm]
{\bf Theorem}\ Suppose that an axially symmetric
electrovacuum spacetime admits a ``piece'' of $\J^+$ in the~sense
that the~Bondi-Sachs coordinates can be introduced in which
the~metric takes the~form (\ref{ds}), (\ref{HRN}), (\ref{dlouhyrozvojmetriky}) and
the~asymptotic form of the~electromagnetic field is given by
(\ref{HRN}), (\ref{dlouhyrozvojmetriky}). If this spacetime admits an additional Killing
vector forming with the~axial Killing vector a two-dimensional Lie
algebra (both Killing vectors need not be hypersurface orthogonal)
then the~additional Killing vector has asymptotically
the~form (\ref{Bbotr}). For $k=0$ it generates asymptotically translations
along the~$z$ or $t$-axis or their combination and then
the~Weyl tensor does not have  radiative components; 
it may contain an infinite thin cosmic string along the~$z$-axis.
For $k\not= 0$ it is the~boost Killing field
and the~spacetime is radiative.

\subsection{General structure of boost-rotation symmetric spacetimes}
\label{secgen}

The~theorem mentioned in the~previous section shows that boost-rotation symmetric spacetimes
play a unique role among all locally asymptotically flat axially symmetric
electrovacuum spacetimes with a further symmetry as they are the~only ones that are
radiative.
Moreover there is a number of  boost-rotation symmetric
solutions known explicitly, representing the~fields of ``uniformly accelerated  sources''
-- singularities of the~Curzon-Chazy type and of all the~other Weyl types or black holes.

Let us now summarize general properties and the~global behaviour
of boost-rotation symmetric spacetimes  from the~geometrical point of view.
The reader may find the following definitions and statements in the detailed work \cite{BicSchPRD} (see also \cite{ujibiLes,Bipreh,BiGal}) where the~Killing vectors are assumed
to be hypersurface orthogonal.

\begin{figure}
\begin{center}
\includegraphics*[height=7cm]{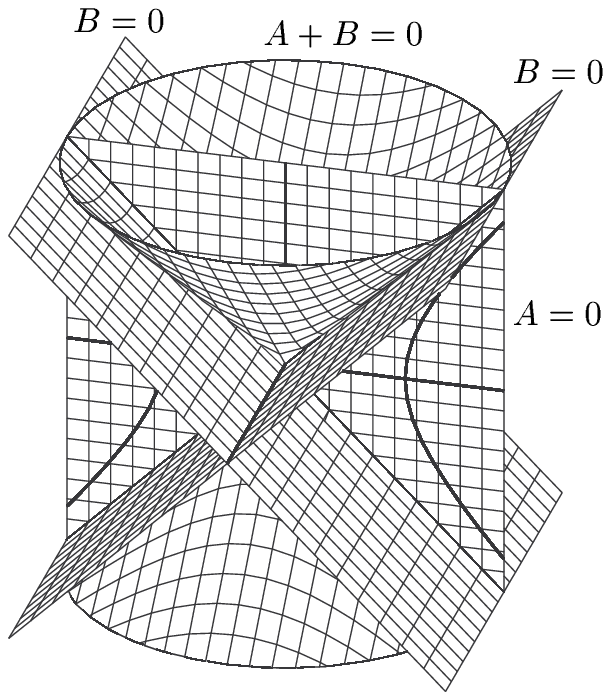}
\includegraphics*[height=200pt]{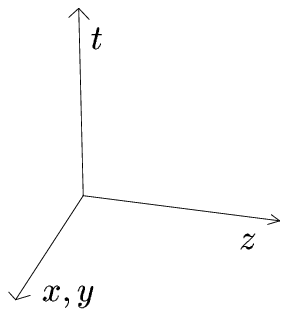}
\end{center}
\caption{The~general structure of  boost-rotation symmetric spacetimes:\protect\\
the~axis -- $A=0$,\protect\\
the~roof -- $B=0$,\protect\\
the~null cone of the~origin -- $A+B=0$.} 
\label{pickuzely}
\end{figure}

Boost-rotation symmetric spacetimes have two isometries generated by two Killing vectors
-- the~axial Killing vector~$\x$ and the~boost Killing vector $\e$.
To gain better insight into curved boost-rotation symmetric spacetimes
let us first consider the~Minkowski spacetime where these two Killing vectors 
and their norms have the~form
\BEA
\mbox{the~axial Killing vector: }\x&=&x\frac{\der}{\der y}-y\frac{\der}{\der x}\ ,
\mm A=-\x_\a\x^\a=x^2+y^2\geq 0\ ,\nn\\
\mbox{the~boost Killing vector: }\e&=&z\frac{\der}{\der t}+t\frac{\der}{\der z}\ ,
\mm B=\ \e_\a\e^\a=z^2-t^2\mm\mbox{arbitrary}\ .\label{KillMink}
\EEA
The~axial Killing vector is spacelike everywhere and vanishes on the~{\it{axis}} \mbox{$A=0$}
containing points fixed by the~rotation. 
The~boost Killing vector is null on two hyperplanes \mbox{$B=0$} -- 
the~{\it{roof}}, spacelike ``above the~roof'' for $B<0$, where ``almost all''
null geodesics get in, and timelike ``below the~roof'' for $B>0$. 
See Fig.~\ref{pickuzely}, where also
the~null cone of the~origin $A+B=0$ is plotted that is fixed under
the~boost-rotation group and its vertex is the~fixed point of the~group
and two integral curves of the~timelike
boost Killing vector (hyperbolas) are indicated that may correspond to worldlines of particles uniformly accelerated 
in opposite directions. Notice that such particles move independently since their worldlines are separated
by two null hypersurfaces (the~roof).
One obtains a similar picture for a curved boost-rotation symmetric spacetime.

Since we wish to examine especially the~radiative properties, let us
concentrate on  the~region above the~roof ($B<0$). We  transform the~Minkowski metric 
first to coordinates adapted to the~boost-rotation symmetry 
$\{ b,\ \r ,\ \f ,\ \chi \}$
\BE
x=\r\cos\f\ ,\mm y=\r\sin\f\ ,\mm z=b\sinh\chi\ ,\mm t=\pm\ b\cosh\chi\label{sourxyzt} 
\EE
with $b\geq 0$, $\chi\in\R$, and further to null coordinates $u$, $v$
\BE
u=b-\r\ ,\mm v=b+\r\ .\label{souruv}
\EE
The~metric then takes the~form
\BE
{\rm d}s^2={\rm d}u{\rm d}v-\ctvrt (v^2-u^2)\lvhz \frac{v-u}{v+u}\ {\rm d}\f^2
+\frac{v+u}{v-u}\ {\rm d}\chi^2\pvhz\ ,\label{dsMinkuv}
\EE
where $v=u$ is the~axis, $v=-u$ the~roof, and the~lines $u$, $\f$, $\chi$ being constant, 
$v$ changing are null geodesics ending at null infinity $\J^+$ for $v\msip +\infty$.
The~Killing vectors (\ref{KillMink}) are $\x=\der /\der\f$ and $\e=\der /\der\chi$.

Inside the~radiative part of the~boost-rotation symmetric spacetime the~boost Killing vector
is spacelike. Thus one cannot distinguish between a small region of a curved boost-rotation symmetric 
spacetime  and a small region of a curved cylindrically symmetric spacetime
(one can transform a boost-rotation symmetric metric above the~roof locally to
the~cylindrically symmetric Einstein-Rosen metric \cite{EinstRos,Kramer}). 
However, these two spacetimes have essentially different asymptotical properties.

The~null coordinates $u$, $v$ are related to  
canonical coordinates $U$, $V$ ($V=\pul v^2$  
and $U=\pul u^2\ {\rm sign}\ u$)
that can be used to distinguish between cylindrical
and boost-rotation symmetry:\\[2mm] 
{\bf{Proposition}} The~metric of a curved spacetime admitting two spacelike,
hypersurface orthogonal, commuting Killing vectors 
$\der /\der{\bar x}$ and $\der /\der{\bar y}$,
and satisfying the~vacuum Einstein equations can be, in a region where $W_{,\a} W^{,\a}$
does not change the~sign, $W$ being the~``volume element'' of the~group orbits,
transformed into one of the~following forms:
\BE
{\rm d}s^2={\rm e}^\Lambda {\rm d}U{\rm d}V-W[{\rm e}^{-\Psi}{\rm d}{\bar x}^2 
+{\rm e}^\Psi {\rm d}{\bar y}^2 
]\ ,
\EE
where $\Lambda$, $\Psi$ are functions of $U$, $V$, and $W>0$ is such that
\BEA
a) &&\  W=\pul (V+U)\ \mbox{ if }\ W,_\a \mbox{ is timelike},\nn\\
b) &&\ W=\pul (V-U)\ \mbox{ if }\ W,_\a \mbox{ is spacelike},\nn\\
c) &&\ \mbox{spacetimes with}\ W,_\a\ \mbox{null or}\ W=\mbox{const have more
symmetries}.\nn
\EEA
The~coordinates $U$ and $V$ are determined
uniquely up to translations.\\[1mm]

In a boost-rotation symmetric spacetime in radiative regions, $W,_\a$ is spacelike
inside the~null cone of the~origin and timelike outside the~null cone, whereas 
$W,_\a$ is spacelike everywhere in a cylindrically symmetric spacetime.
The~Killing vectors $\der /\der {\bar x}$ and $\der /\der {\bar y}$  are the~axial 
and the~boost Killing vector respectively.

The~metric (\ref{dsMinkuv}) can be generalized for a curved spacetime:\\[2mm]
{\bf{Definition 1}} A spacetime admitting two spacelike, hypersurface orthogonal
Killing vectors is called ``boost-rotation symmetric'' if in canonical coordinates
$\pul v^2$, $\pul u^2$ (resp. $-\pul u^2$ if  $u<0$) the~metric has the~form
\BE
{\rm d}s^2={\rm e}^\l\ {\rm d}u{\rm d}v-\ctvrt(v^2-u^2)\lvhz
\frac{v-u}{v+u}\ {\rm e}^{-\m}\ {\rm d}\f^2+\frac{v+u}{v-u}\ {\rm e}^\m\ {\rm d}\chi^2\pvhz\ ,
\label{dsuv}
\EE
$\f\in\langle 0,\ 2\p)$, $\chi\in\R$, the~functions $\l(u,\ v)$, $\m(u,\ v)$
are defined for $v\in (0<v_0,\ +\infty)$, $u\in(u_0,\ u_1)$, $u<v$, $u\not= -v$, and
\BE
\lim_{{v\to \infty} \atop {u\ \mbox{{\tiny{fixed}}}}}
         \l (u,\ v)=\l_0(u)\ ,\mm
\lim_{{v\to \infty} \atop {u\ \mbox{{\tiny{fixed}}}}}\m (u,\ v)=
                \k=\mbox{const}\ .\label{limlmu}
\EE
\\[1mm]

This weaker asymptotic condition is compatible with the~asymptotic flatness --
boost-rotation symmetric spacetimes defined here admit a local $\J^+$. A stronger
condition $\l\msip 0$, $\m\msip 0$ as $v\msip +\infty$ would imply a flat spacetime
everywhere.

The~boost-rotation symmetric metric defined in Def.~1 (\ref{dsuv}) 
above the~roof, i.e., for $t^2>z^2$, 
can be extended to all values of ``Cartesian-type'' coordinates  $\{ t,\ x,\ y,\ z\}$ 
by transformations  (\ref{sourxyzt}) and (\ref{souruv}):\\[2mm]
{\bf{Definition 2}} The~boost-rotation symmetric vacuum solutions have the~metric
\BEA
{\rm d}s^2&=&-(x^2+y^2)^{-1}
[{\rm e}^\l (x{\rm d}x+y{\rm d}y)^2
          +{\rm e}^{-\m}(x{\rm d}y-y{\rm d}x)^2] 
-(z^2-t^2)^{-1} [{\rm e}^\l(t{\rm d}t-z{\rm d}z)^2
      -{\rm e}^\m(t{\rm d}z-z{\rm d}t)^2]\nn\\
     &=&-(x^2+y^2)^{-1}
[({\rm e}^\l x^2+{\rm e}^{-\m}y^2){\rm d}x^2 
           +2xy({\rm e}^\l-{\rm e}^{-\m}){\rm d}x\ {\rm d}y
           +({\rm e}^\l y^2+{\rm e}^{-\m}x^2){\rm d}y^2]\nn\\
      &&-(z^2-t^2)^{-1}[({\rm e}^\l z^2-{\rm e}^\m t^2){\rm d}z^2 
           -2zt({\rm e}^\l-{\rm e}^\m){\rm d}z\ {\rm d}t
               -({\rm e}^\m z^2-{\rm e}^\l t^2){\rm d}t^2]\ 
              \label{dstxyz}
\EEA
with $(t,\ x,\ y,\ z)\in\R^4$, $x^2+y^2>0$, $z^2-t^2\not= 0$, and
\BE
\m=\m(A=x^2+y^2,\ B=z^2-t^2)\ ,\mm\l=\l(A=x^2+y^2,\ B=z^2-t^2)\ .\label{mulAB}
\EE
The~function $\m$, determined up to an additive constant, is
an analytic, asymptotically regular solution of the~flat-space wave equation
\BE
\Box\m=A\m,_{AA}+B\m,_{BB}+\m,_A+\m,_B=0\label{rcemu}
\EE
 except for the~regions where sources uniformly
accelerated with respect to the~Minkowskian background and defining $\m$ occur.
The~function $\l$, determined up to an additive constant, 
is the~solution of the~field equations
\BEA
(A+B)\l,_A&=&B(A{\m,_A}^2-B{\m,_B}^2+2A\m,_A\m,_B)+(A-B)\m,_A-2B\m,_B\ ,\label{rcel1}\\
(A+B)\l,_B&=&A(B{\m,_B}^2-A{\m,_A}^2+2B\m,_A\m,_B)+(A-B)\m,_B+2A\m,_A\ \label{rcel2}
\EEA
and it is analytic in all regions where $\m$
is analytic, i.e., both are analytic even on the~roof and everywhere on the~axis
outside the~sources and nodal (conical) singularities which cause the~particles
to accelerate. The~equation (\ref{rcemu}) is an integrability condition for equations (\ref{rcel1}) 
and (\ref{rcel2}).\\[1mm]

One can see  that $\m$ and $\l$ (\ref{mulAB}) in canonical coordinates satisfy the~boundary conditions 
required in Def.~1
and the~metric admits the~axial and boost Killing vectors which have the~same form
as in the~Minkowski spacetime (\ref{KillMink}). Moreover, the~structure
of group orbits in the~curved boost-rotation symmetric spacetime outside the~sources
or singularities is the~same as the~structure of axial and boost group orbits
in the~Minkowski spacetime. Thus the~boost Killing vector is timelike below the~roof,
$z^2>t^2$, where the~spacetime is  stationary and there is no radiation.
It  is spacelike above the~roof, $t^2>z^2$, where almost all null geodesics end
and where the~leading term of the~Riemann tensor, proportional to $r^{-1}$ (\mbox{$r^2=\r^2+z^2$}),
is non-vanishing and has the~same algebraic structure as in the~case of plane waves.
Consequently 
in  regular regions of $\J$ above the~roof the~gravitational field is radiative.

This definition of vacuum boost-rotation symmetric spacetimes is ``geometrical''
by virtue of an invariant geometrical meaning of ``Minkowskian'' coordinates
$\{ t,\ x,\ y,\ z\}$ due to their relation to the~canonical coordinates.

Using polar coordinates $\rho$, $\f$ instead of $x$, $y$, the~metric (\ref{dstxyz})
becomes
\BE
{\rm ds}^2 = -{\rm e}^{\lambda} {\rm d} \rho^2 - \rho^2 {\rm e}^{-\mu} {\rm d}\f^2 
- \frac{1}{z^2-t^2} \left[ ({\rm e}^{\lambda} z^2 - {\rm e}^{\mu} t^2 ) {\rm d} z^2
 -   2zt ({\rm e}^{\lambda} - {\rm e}^{\mu}  ) {\rm d} z\  {\rm d} t 
+     ({\rm e}^{\lambda} t^2 - {\rm e}^{\mu} z^2 ) {\rm d} t^2  \right]    \ .   
\label{BStvar}
\EE

Since both the~functions $\m$ and $\l$  are determined up to an additive constant, we
can make the~roof regular, i.e., 
\BE
\m(A,\ 0)=\l(A,\ 0) \label{regstrecha}
\EE
is satisfied,  by choosing the~constants such that 
\BE
\m(0,0)=\l(0,0)\ \label{regroofpoc}
\EE
is fulfilled as the~Einstein equations (\ref{rcemu})--(\ref{rcel2})  then imply (\ref{regstrecha}).
As the~sources are located on the~axis, we  cannot make the~whole axis regular but
one can add the~same constant to both $\m$ and $\l$ to make parts of the~axis regular, i.e., 
to satisfy
\BE
\m(0,\ B)+\l(0,\ B)=0\ ,\label{regaxis}
\EE
and thus arrange
the~distribution of nodal (conical) singularities -- ``strings'' -- along the~$z$-axis.
One can make regular either the~spatial infinity
(then the~string is between  symmetrically located sources on the~$z$-axis)
or the~axis between sources moving with opposite accelerations
(then the~string tends to infinity along \mbox{the~$z$-axis} and thus propagates along
null infinity to time infinity). 
Such solutions are asymptotically flat in the~sense that they admit only local ${\cal J}$ 
(two generators are singular, or missing
so that ${\cal J}^\pm \not\cong S^2\times{\cal R}$).
See Sec. \ref{sec3} for particular examples.

There are also solutions  where no nodal singularity is necessary to cause the~acceleration
of the~sources. Thus both the~roof and axis (except for points where particles occur) 
are regular 
and the~regularity conditions of the~roof  (\ref{regroofpoc}) and axis (\ref{regaxis}) 
at the~origin  imply $\m(0,0)=\l(0,0)=0$.
Both spacelike and timelike infinities are then regular. 
Null infinity $\J^+$ and $\J^-$ is regular except for four points, where 
the~worldlines of sources ``start'' and ``end'' ($t=\pm z$, $z\msip\pm\infty$) --
the~fixed points of the~boost-rotation symmetry, i.e.,
the~solution admits global null infinity in the~sense of 
topology ${\cal J}^\pm \cong S^2\times{\cal R}$
though two generators are not complete.  
It can be shown that 
arbitrarily strong boost-rotation symmetric initial data
can be prescribed on a hyperboloidal hypersurface above the~roof leading 
to a complete smooth future null and timelike infinities. Here weak-field initial data are not 
necessary as in the~proof of existence of general asymptotically flat
radiative solutions (the~work of Friedrich \cite{Friedrich}, Cutler and Wald \cite{CW}, Christodoulou and Klainerman \cite{ChK}).
These exceptional cases can be represented by ``self-accelerating'' particles
due to their multipole structure (an infinite number of these solutions is
known in an explicit form \cite{jibiHS}) or by several particles
distributed symmetrically along \mbox{the~$z$-axis} where negative masses have to be
introduced (for example \cite{BSZ}). 

``Generalized'' boost-rotation symmetric solutions also do not contain singularities 
causing the~acceleration and even no negative masses. 
They represent accelerated particles in  external fields.  
The~generalized C-metric was obtained by a certain procedure by Ernst \cite{Ernst78}, 
the~generalized Bonnor-Swaminarayan solution was first constructed 
analogously and  then in a more physical way as a limit of the~asymptotically flat
boost-rotation symmetric solution representing two accelerated pairs when the~``outer'' 
particle in each pair, attached to a string extending to infinity, 
goes to infinity with increasing mass parameter \cite{BHSI}.
These solutions are the~best rigorous  known today examples 
 of the~motion of relativistic bodies, e.g., black holes in ``external fields''.
However,  they are not asymptotically flat and thus it is difficult to analyze 
their radiative properties. In a limit of a weak external field,
there are regions where radiation properties might be investigated since
spacetimes are approximately flat there. Recently Hawking and Ross \cite{Hawk} used
the~generalized C-metric within the~framework of quantum gravity. 
There exists also an electromagnetic generalization
\cite{Ernst76}, where the~charged black hole is accelerated by an external electric field 
(the~same solution may be achieved by interacting electromagnetic and gravitational
perturbations in a weak external limit \cite{jibinaboje}).

As was mentioned at the~beginning, asymptotically flat
boost-rotation symmetric spacetimes are the~only non-trivial explicit exact radiative 
solutions representing moving objects and an analysis of their radiative 
properties may be helpful to understand physically more realistic situations.
These solutions are the~most realistic, although they contain nodal singularities
or negative masses and their ADM mass vanishes,
and thus can not serve as an example 
of a general spacetime with a positive ADM mass.

\section{Examples in electromagnetism and gravity}
\label{sec3}

  \subsection{Uniformly accelerated electromagnetic multipoles}
\label{secelmg}

In this section  we study boost-rotation symmetric solutions 
of the~Maxwell equations to gain better understanding
of solutions with the~same symmetries in general theory of relativity.
The~simplest case  was studied first by Born \cite{Born}. It represents a field of two   particles with opposite charges $\pm e$ symmetrically located and uniformly accelerated
with a uniform acceleration $\pm\alpha^{-1}$, $\a>0$, along the~$z$-axis of
cylindrical coordinates $\{ t,\rho,\phi,z\}$ in the~Minkowski 
spacetime (see, e.g., \cite{Rohrlich}).
They move independently of each other as they are separated by a null hypersurface, the~roof
(see Subsec.~\ref{secgen}).
Their worldlines are  hyperbolas, integral curves of the~timelike boost Killing vector,
\BE
\rho=0, \quad  z_{(\pm e)}=\pm\sqrt{\alpha^2+t^2}\ . \label{svetocar}
\EE
A detailed analysis \cite{Rohrlich,bicakBS} shows that the~field can be interpreted as 
either the~purely retarded field from the~charge $+e$ in the~region $z+t>0$ and
the~purely advanced field from the~charge $-e$ in the~region $z+t<0$, or as
$\pul$(advanced $+$ retarded) fields from both charges. However, the~field is 
purely retarded in the~region $t>|z|$  and purely advanced in  $t<-|z|$.
The~same interpretation may be used for a field of 
uniformly accelerated electric   \cite{Muschall}
and  magnetic  \cite{VPhD,AVWDS} multipoles.

The~field of  uniformly accelerated electric multipoles ($2^l-$poles) $\pm{\mathtt{e}}_l$
moving along hyperbolas  (\ref{svetocar}) reads (see \cite{Muschall})
\BEA
E^{(\rho)}_{(l)} &=&  \enspace  \ 8\  \frac{{\mathtt{e}}_l (2l-1)!!}{l!^2}\  \rho z\  \partial_\alpha^{(l)} 
\left ({\frac {{\alpha}^{2} }{{\z}^{3}}}\right ) \ ,  \quad \nonumber \\
E^{(z)}_{(l)} &=& -4\ \frac{{\mathtt{e}}_l (2l-1)!!}{l!^2}\  \partial_\alpha^{(l)} 
\left ({\frac {{\alpha}^{2} \left ({\alpha}^{2}+{t}^{2}+{\rho}^{2}-{z}^{2}\right )}
{{\z}^{3} }}\right )  \ ,   \label{Felmulti} \\
B^{(\phi)}_{(l)} &=&    \enspace \ 8\ \frac{{\mathtt{e}}_l (2l-1)!!}{l!^2}\ \rho t \ 
\partial_\alpha^{(l)} \left ({\frac {{\alpha}^{2}}{{\z}^{3}}}\right ) , \nonumber
\EEA
while the~field corresponding to  uniformly accelerated magnetic multipoles $\pm{\mathtt{m}}_l$
has the~form (see \cite{VPhD,AVWDS})
\BEA
B^{(\rho)}_{(l)}&=& \enspace  \ 8\  \frac{{\mathtt{m}}_l (2l-1)!!}{l!^2}\ \rho z \ 
\partial_\alpha^{(l)} \left ({\frac {{\alpha}^{2}}{{\z}^{3}}}\right ) \ ,  \nonumber \\
B^{(z)}_{(l)} &=& - 4\ \frac{{\mathtt{m}}_l (2l-1)!!}{l!^2}\  
\partial_\alpha^{(l)} \left ({\frac {{\alpha}^{2} 
\left ({\alpha}^{2}+{t}^{2}+{\rho}^{2}-{z}^{2}\right )}{{\z}^{3}}}\right )  \ ,  \label{Fmgmulti} \\
E^{(\phi)}_{(l)}&=& -  8\ \frac{{\mathtt{m}}_l (2l-1)!!}{l!^2}\ \rho t \ 
\partial_\alpha^{(l)} \left ({\frac {{\alpha}^{2}}{{\z}^{3}}}\right ) , \nonumber
\EEA
where
\BE
\z=\sqrt{4\rho^2\alpha^2+(\alpha^2+ t^2-z^2-\rho^2)^2} \ .
\EE
The~other field components vanish.
It is obvious that the~field given by (\ref{Fmgmulti}) can be obtained from that given
by (\ref{Felmulti})
by a simple transformation
\BE
\vec B \rightarrow -\vec E, \quad \vec E \rightarrow \vec B \ ,
\quad {\mathtt{e}}_l \rightarrow {\mathtt{m}}_l \ ,
\label{dual1}
\EE
which  is a special case of the~duality symmetry.

\begin{figure}
\begin{center}
\includegraphics*[height=5cm]{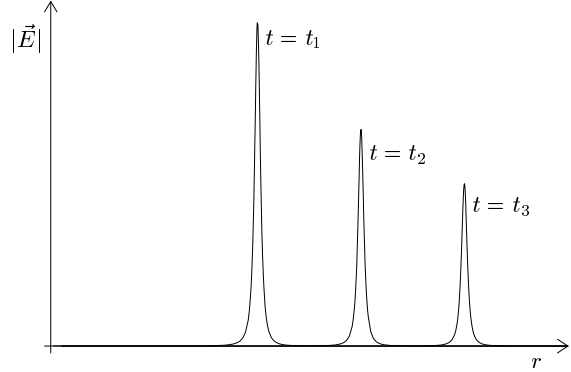}
\end{center}
\caption{Pulse generated by a uniformly accelerated electric charge.}
\label{fig:pulse}
\end{figure}

Now let us analyze these solutions from the~viewpoint of radiation.
The~Born solution, the~simplest example of  (\ref{Felmulti}) 
representing the~field of uniformly accelerated electric monopoles, was
studied in some basic works by Pauli \cite{Pauli} and von Laue \cite{Laue}.
They considered the~field
as non-radiative as do some authors even now
\cite{Singal} (see also comments in \cite{Parrot,HarSok}).
However, this  is in a contradiction with the~statement that 
an accelerated charge radiates energy with the~rate $(2/3) e^2 {\dot v}^2$. 
Performing a series expansion in $r^{-1}$ $(r^2=\rho^2+z^2)$ with 
time $t$ fixed, the~asymptotic behaviour of the~Born field is 
$\vec E \sim r^{-4}$, $\vec B \sim r^{-5}$
and thus  the~Poynting vector $\vec S \sim r^{-9}$. In Fig. \ref{fig:pulse} we
 see that the~quantities determining the~field have 
the~character of a pulse and it is therefore understandable  why 
the~Poynting vector $\vec S $ is non-radiative when going
to spatial infinity and therefore passing through a pulse (see \cite{bicakBS}). In 
the~next part we will see that $\vec S \sim r^{-2}$ when travelling
with the~pulse with the~velocity of light; then all fields
(\ref{Felmulti}) and (\ref{Fmgmulti})  have a radiative character.
To examine the~radiative properties of these solutions,
we  express the~field components   (\ref{Felmulti}) and (\ref{Fmgmulti}) in terms of spherical coordinates
$(r,\theta,\f)$ and of the~retarded time of the~origin $u=t-r$
and expand them in $r^{-1}$ with $u$, $\theta$, $\f$ fixed (see \cite{Muschall,VPhD,AVWDS}):
\BEA
B^{(\rho)} &=& \enspace  \ {\mathtt{m}}_l G_l \sin \theta \cos \theta\ \frac{1}{r}
+{\cal O}(r^{-2})   \ ,
\nonumber \\
B^{(z)} &=& -{\mathtt{m}}_l G_l \sin^2 \theta\ \frac{1}{r} 
+{\cal O}(r^{-2})    \ ,   \label{leading} \\
E^{(\phi)} &=& -{\mathtt{m}}_l G_l \sin \theta\  \frac{1}{r} 
+{\cal O}(r^{-2})  \ , \nonumber
\EEA
where
\BE
G_l=\frac{(2l-1)!!}{l!^2} \ 
\partial_\alpha^{(l)} \left (     
\frac{\alpha^2}{(u^2+\alpha^2 \sin^2 \theta )^{3/2}}\right)\  \label{Gl}
\EE
and similarly for electric multipoles with respect to (\ref{dual1}).
Considering  a particle with an arbitrary structure of electric and
magnetic multipoles, the~leading term of the~radial component of the~Poynting vector
reads
\BE
S_r=\frac{\sin^2 \theta}{4 \pi r^2} \sum_{l,l'}
({\mathtt{e}}_l {\mathtt{e}}_{l'}+{\mathtt{m}}_l {\mathtt{m}}_{l'}) G_l G_{l'} +
{\cal O}(r^{-3}) \ .
\EE 
Introducing the~true retarded time $u^*$ of the~particle by
\BE
t-u^*=|\vec r - \vec r^{\, *}| \ ,
\EE
where $(t,\vec r)$ and $(u^*,\vec r^{\, *})$ are the~coordinates of an observation 
and emission event, respectively,  we can express
\BE
u=u^*-\sqrt{{u^*}^2+\alpha^2} \cos \theta + {\cal O}(r^{-1})
\label{defu} \ .
\EE
The~radial flux emitted at $u^*=0$ thus takes the~form
\BE
S_r=\frac{\sin^2 \theta}{4 \pi r^2} \sum_{l,l'}
 \frac{(2l-1)!! (2l'-1)!!({\mathtt{e}}_l {\mathtt{e}}_{l'}+{\mathtt{m}}_l {\mathtt{m}}_{l'})}{l!^2
 l'!^2}  \frac{V_l V_{l'}}{\alpha^{l+l'+2}} \ ,
\EE
where $V_l$ can be obtained by substituting (\ref{defu}) and $u^*=0$
into
\BDM
\alpha^{l+1} \, \partial_\alpha^{(l)} \left ( 
\frac{\alpha^2}{(u^2+\alpha^2 \sin^2 \theta )^{3/2}}\right) \ ;
\EDM
thus $V_0=1$, $V_1=3 \cos^2 \theta -1$, $V_2=15 \cos^4 \theta-15 \cos^2
\theta+2$, $\dots$. \\
The~total power radiated out,  $R=2 \pi \int_{0}^{\pi} {\lim\limits_{r \to \infty}}
(r^2 S_r) \sin \theta \, {\rm d} \theta$, is then as follows 
\BE
R=\sum_{l,l'} R^{(l,l')} \frac{{\mathtt{e}}_l {\mathtt{e}}_{l'}+{\mathtt{m}}_l 
{\mathtt{m}}_{l'}}{{\alpha^{l+l'+2}}} \label{power} \ ,
\EE
where $R^{(l,l')}$ are introduced by  (see (24) in \cite{Muschall})
\BE
R^{(l,l')}=\frac{(2l-1)!! (2l'-1)!!}{2l!^2 l'!^2} \int_{0}^\pi V_l
V_{l'} \sin^3 \theta \, {\rm d} \theta \ .
\EE
This expression for the~total radiated power (\ref{power})  reduces to (23) in \cite{Muschall}
when all ${\mathtt{m}}_l=0$. 
As in \cite{Muschall} we have
 $R^{(0,0)}=2/3$, $R^{(0,1)}=-4/15$, $R^{(1,1)}=8/21$, etc. Due to the~boost
 symmetry, the~particles radiate out energy (measured in an inertial frame in which
the~particle is at rest at the~emission event)
with a constant rate which is
 independent of $u^*$ and thus the~same as at the~turning point $u^*=0$. 
See \cite{bicakBS} for the~detailed discussion.

From (\ref{dlouhyrozvojmetriky}), (\ref{boost-X}), and (\ref{boost-Y}) 
it follows that radiative components 
of a general boost-rotation symmetric 
electromagnetic field tensor $\Fmn$ in the~coordinates $u$, $r$, $\theta$, $\phi$
are of the~form
\BEA
F_{u\theta} &=& X=\frac{\E(w)}{u^2}
\ , \nonumber \\
F_{u\phi} &=& Y\sin\theta=\frac{\tilde{\BB}(w)\sn}{u^2}
\ ,\mm {\rm where} \ \  w=\frac{\sin\theta}{u} \ . 
\EEA
Here  $X$ and $Y$ correspond to the~electromagnetic news functions of the
system (see Subsubsec.~\ref{secbondi} or \cite{burg,ajajibi}).
Notice that these radiative components are proportional to $r^0$ rather then to $r^{-1}$
due to the~choice of coordinates.
For a uniformly accelerated
electric monopole (i.e., the~Born solution) we obtain
\BEA
\E (w)&=&\frac{{\mathtt{e}}_0 \alpha^2 w}{(1+\alpha^2 w^2)^{3/2}} \mm\mbox{for}\ u> 0\ ,\mm
\E (w)=-\frac{{\mathtt{e}}_0 \alpha^2 w}{(1+\alpha^2 w^2)^{3/2}} \mm\mbox{for}\ u<0\ , \nonumber \\
\tilde{\BB}(w)&=&0. 
\EEA
For a uniformly accelerated
electric dipole  we get
\BEA
\E (w)&=&\frac{{\mathtt{e}}_1 \alpha w(2-\alpha^2 w^2)}{(1+\alpha^2 w^2)^{5/2}}  \mm\mbox{for}\ u> 0\ ,\mm
\E(w)=-\frac{{\mathtt{e}}_1 \alpha w(2-\alpha^2 w^2)}{(1+\alpha^2 w^2)^{5/2}}  \mm\mbox{for}\ u< 0\ , \nonumber \\
\tilde{\BB}(w)&=&0 \ , 
\EEA
 whereas for a magnetic dipole  one finds
\BEA
\E (w)&=&0 \ , \nonumber \\
\tilde{\BB}(w)&=&- \frac{{\mathtt{m}}_1 \alpha w(2-\alpha^2 w^2)}{(1+\alpha^2 w^2)^{5/2}} \mm\mbox{for}\ u> 0\ ,\mm
\tilde{\BB}(w)= \frac{{\mathtt{m}}_1 \alpha w(2-\alpha^2 w^2)}{(1+\alpha^2 w^2)^{5/2}} \mm\mbox{for}\ u<0\ . 
\EEA
For a general electric $2^l$-pole the~news functions turn out to be
\BEA
\E (w)&=&{\mathtt{e}}_l (G_l u^3)w \ , \nonumber \\
\tilde{\BB} (w)&=&0 \ , 
\EEA
while for a magnetic $2^l$-pole
\BEA
\E (w)&=&0 \ , \nonumber \\
\tilde{\BB} (w)&=&-{\mathtt{m}}_l (G_l u^3)w  
\EEA
with $G_l$ given by (\ref{Gl}).

One can check that the~total electric and magnetic charges (\ref{zachnab}) for the~general 
electric and magnetic multipoles vanish,
\BE
\int\limits_{0}^{\p} (\E\cs /u) {\rm d}\th=0\ ,\mm
\int\limits_{0}^{\p} (\tilde{\BB}\cs /u) {\rm d}\th =0\ .
\EE


  \subsection{The~Bonnor-Swaminarayan solutions}
\label{secBS}


In 1957 Bondi \cite{bondineg} proved the~existence of an axially symmetric solution of the~Einstein equations
corresponding to two pairs of particles having equal masses but of opposite sign 
uniformly accelerated in opposite directions.  Seven years later Bonnor and Swaminarayan 
constructed such a solution explicitly \cite{BSZ}. Particles in this solution
-- point singularities of Curzon-Chazy type  (see \cite{Kramer}) --  move freely and
the~acceleration force is caused by mutual gravitational interaction. 
They obtained also other similar metrics, some of them representing only positive masses.
However, particles considered there do not move freely, line singularities of a conical type, 
interpreted as nodes, rods, struts or stresses, always appear.

Since Bonnor-Swaminarayan  solutions (BS-solutions) 
are boost-rotation symmetric, their metrics have,
in cylindrical coordinates $t$, $\rho$, $z$, and $\phi$, the~form 
(\ref{BStvar}),
in which the~functions entering the~metric read 
\BEA
\mu&=&-\frac{2a_1}{R_1}-\frac{2a_2}{R_2}
     +\frac{2a_1}{h_1}+\frac{2a_2}{h_2}+\ln k\ ,\nonumber\\
\lambda&=&\frac{a_1 a_2}{(h_1-h_2)^2}f
     -\rho^2(z^2-t^2)\lvvkz\frac{a_1^2}{R_1^4}
           +\frac{a_2^2}{R_2^4}\pvvkz
     +\frac{2a_1 R}{h_1 R_1}+\frac{2a_2 R}{h_2 R_2}+\ln k\ ,\nonumber\\
R &=&\pul (\rho^2+z^2-t^2)\ ,\label{BSobecne}\\
R_i&=&\sqrt{(R-h_i)^2+2\rho^2 h_i}\ ,\mm i=1,2\ ,\nonumber\\
f  &=&\frac{4}{R_1 R_2}[
\rho^2(z^2-t^2)+(R-\rho^2-h_1)(R-\rho^2-h_2)-R_1R_2]\ ,\nonumber
\EEA
where $a_1$, $a_2$, $h_1>0$, $h_2>0$, $k>0$ are constants.

The~whole class of BS-solutions  
describes the~gravitational field of a finite number of monopole Curzon-Chazy particles (CC-particles)
uniformly accelerated in opposite directions where  the~acceleration force
is caused  by a gravitational interaction among particles or by
nodal singularities. Let us now study two examples separately.

%
%
\subsubsection{{{\mbox{The~Bonnor-Swaminarayan}
 \mbox{solution} \mbox{representing}
two pairs of  \mbox{freely} \mbox{moving}
\mbox{CC-particles} \mbox{without} a \mbox{nodal} 
 \mbox{singularity}}}}

The~simplest case without a nodal singularity is the~BS-solution, considered first by Bondi,
containing two pairs of freely falling particles: there is one particle with positive
and one with negative mass in each pair.
In this case the~constant parameters in (\ref{BSobecne}) are given by
\BE
a_1=\frac{(h_1-h_2)^2}{2h_2}\ ,\mm
a_2=-\frac{(h_1-h_2)^2}{2h_1}\ ,\mm
k  =1\ .
\EE

To examine the~radiative properties and to find the~news function
of this solution, we shall  transform the~metric at
first to spherical flat-space coordinates
$\{R,\ \vartheta,\ \phi\}$ by
$\rho=R \sin\vartheta$, $z=R\cos\vartheta$, $\phi=\phi$,
and flat-space retarded time $U=t-R$ (see \cite{bicakBS}) 
and then find a transformation to the~Bondi coordinates $u$, $r$,
$\theta$, and $\phi$, where
the~metric has the~Bondi form (see (\ref{ds}) 
 or  \cite{bondi} and (2), (4) in \cite{bicakBS}).

For $h_2=h$, $h_1=h+{\tilde \ep}$, ${\tilde \ep}>0$ small, $k=1$ we obtain
a slightly curved spacetime with  the~masses given by
$m^{(1)}={\tilde \ep}^2 / [2h\sqrt{2h+2{\tilde \ep}}]$,
$m^{(2)}=-{\tilde \ep}^2 /[(2h+2{\tilde \ep})\sqrt{2h}]$.
The~relation between the~flat coordinates 
$\{U,\ R,\ \vartheta,\ \phi\}$
and the~Bondi coordinates $\{u,\ r,\ \theta,\ \phi\}$ is then (see \cite{bicakBS})
\BE
U=[ u+{\cal O}({\tilde \ep}^3 )] +{\cal O}(r^{-1})\ ,\mm R=r+{\cal O}(1)\ ,\mm 
\vartheta=\theta+{\cal O}(r^{-1})\ ,\mm \phi=\phi\ .
\EE
The~only non-vanishing news function of this system
in the~Bondi coordinates,  neglecting higher terms in $\ep$, reads
\BEA
c,_u&= & {\tilde \ep}^3\frac{3u\sin^2\theta}{2h(u^2+2h\sin^2\theta)^{5/2}}
             \  \nonumber\\
    &= & {\tilde \ep}^3\frac{3w^2  }{2h(1  +2hw^2 )^{5/2}}\frac{1}{u^2}\mm
         {\rm{for}}\  u> 0\ ,\nn\\
    &= &-{\tilde \ep}^3\frac{3w^2  }{2h(1  +2hw^2 )^{5/2}}\frac{1}{u^2}\mm
         {\rm{for}}\  u<0\ ,\mm {\rm{where}} \mm 
           w=\frac{\sin\theta}{u}\ .\label{cuBSZ}
\EEA
One obtains the~same news function (up to a constant) for 
another solution found in \cite{jibiHS} as was shown in \cite{BicTN}.
It is constructed from the~BS-solution representing two pairs of particles
uniformly accelerated along the~symmetry axis in opposite directions and 
containing one positive and one negative mass connected by a nodal singularity
(this case was not considered explicitly in \cite{BSZ}).
A limiting procedure that brings particles in each pair together simultaneously
increasing their masses leads to a field representing two independent self accelerating
(0,1)-pole CC-particles called {``mo\-no\-di\-pe\-ros''} (the~monopole and dipole
parts are present in the~field function). Except for places where particles occur, the~spacetime
is regular. An octupole radiation pattern was demonstrated  for the~BS-solution representing 
freely moving CC-particles  in \cite{bicakBS} and  for the~field of mo\-no\-di\-pe\-ros in \cite{BicTN}.

\begin{figure}
\begin{center}
\includegraphics*[height=5cm]{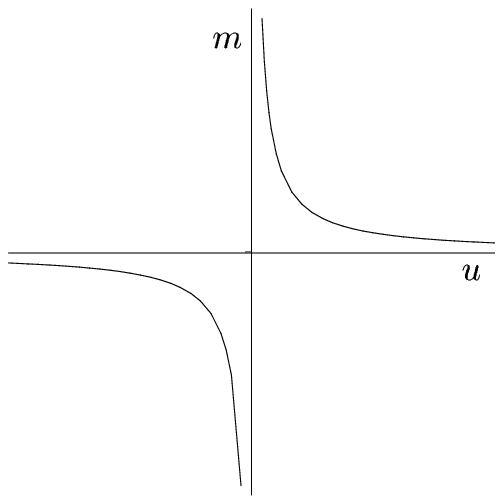}
\hspace{10mm}
\includegraphics*[height=5cm]{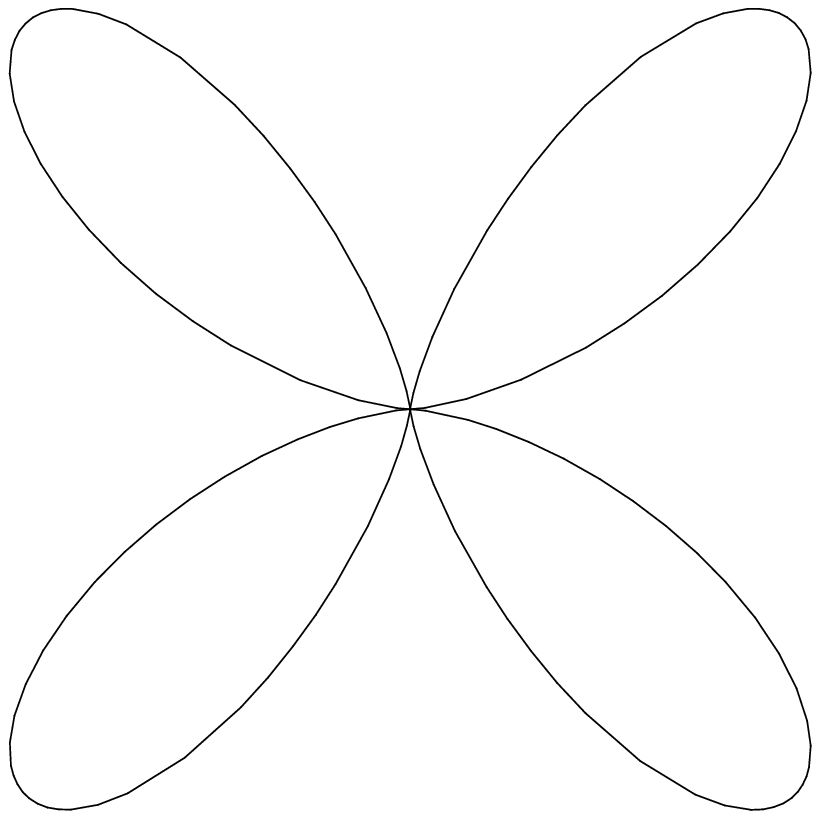}
\end{center}
\caption{Two pairs of freely moving CC-particles:\protect\\
a) the~decrease of the~total Bondi mass at $\J^+$ (\ref{mfree})  and \protect\\
b) the~radiation pattern emitted at the~turning point $u^*=0$.}
\label{pichmotaBSzap}
\end{figure}

Comparing (\ref{cuBSZ}) with the~news function corresponding to
a general  asymptotically flat boost-rotation symmetric solution of the~Einstein equations
(\ref{boost-c}), 
one sees that
\BEA
{\K},_w (w)&=&\ {\tilde \ep}^3\frac{3w^2  }{2h(1  +2hw^2 )^{5/2}}\mm
                  {\rm{for}}\mm u> 0\ ,\nn\\
{\K},_w (w)&=&-{\tilde \ep}^3\frac{3w^2  }{2h(1  +2hw^2 )^{5/2}}\mm
                  {\rm{for}}\mm u<0\ .
\EEA
Then  solving  Eq.~(\ref{boostlambda}) 
for the~function $\l(w)$ and substituting the~result into
Eq.~(\ref{boost-M}), 
we finally integrate Eq.~(\ref{boostm}) 
 to get the~total Bondi mass at $\J^+$
\BEA
m&= &\frac{{\bar {\ep}}^2u^3(5u^4+32u^2h+64h^2)}
              {4h^{7/2}(u^2+2h)^{7/2}} \sqrt{2} 
\times     \lvhz -\ln(u^2)+\ln \left |-2\sqrt{2h}\sqrt{u^2+2h}
             +u^2+4h\right |\pvhz\nn\\
 &\ &+\frac{{\bar \epsilon}^2(u^2+4h)(15u^4+16u^2h+32h^2)}{3h^3u(u^2+2h)^3}\
          ,\mm 
            {\bar \epsilon}=\frac{3{\tilde \ep}^3}{2^6 h}\   .\label{mfree}
\EEA
In Fig. \ref{pichmotaBSzap}a we see that the~Bondi mass $m$ (\ref{mfree}) as a function of $u$ is
everywhere decreasing and thus the~system does radiate gravitational
waves. One can easily verify that the~axis is regular here, i.e.,
$c, _u/ \dsn$ is a  regular function  in the~limit $\sn\ \msip\ 0$. 
Fig.~\ref{pichmotaBSzap}b contains a radiation pattern   
of this solution at the~turning point, i.e., a polar diagram of the~function $c,_u^2(\th,\ u^*=0)$
(see (\ref{defu}) for the~definition of $u^*$) -- 
a gravitational radiation flow to different directions at the~turning point.
\\[2mm]

%
%
\subsubsection{{{\mbox{The~Bonnor-Swaminarayan}
\mbox{solution} \mbox{representing}
two \mbox{CC-particles} 
\mbox{connected} with a \mbox{nodal} \mbox{singularity}}}}

Another example of a boost-rotation symmetric solution is
the~BS-solution (\ref{BStvar}), (\ref{BSobecne}) 
describing  two symmetrically
located CC-particles, uniformly accelerated in opposite
directions and connected with a conical singularity of a finite length
given by the~constants 
\BE
\ln k=-4mA\ ,\mm a_1=\frac{m}{A}\ ,\mm h_1=\frac{1}{2A^2}\ ,
\mm a_2=h_2=0\ .
\EE
As in the~previous case we transform the~metric 
first to flat-space coordinates  $\{U,\ R,\ \vartheta,\ \phi\}$
and then to  the~Bondi coordinates  $\{u,\ r,\ \theta,\ \phi\}$
that are connected in limits of  small mass $mA\ll 1$ and $r\msip\infty$ by
\BE
U=[ u+{\cal O}(mA)] +{\cal O}(r^{-1})\ ,\mm R=r+{\cal O}(1)\ ,\mm 
\vartheta=\theta+{\cal O}(r^{-1})\ ,\mm \phi=\phi\ .
\EE
Then the~news function in the~Bondi coordinates reads
(see (28) in \cite{jibistruna})
\BEA
c,_u&= & \frac{2mA}{\dsn}
        +\frac{mu}{A(u^2+\dsn/A^2)^{3/2}}
        +\frac{2mAu}{\dsn\sqrt{u^2+\dsn/A^2}} \nn\\
    &= & \lvvhz \frac{2mA}{w^2}
        +\frac{m}{A(1+w^2/A^2)^{3/2}}
        +\frac{2mA}{w^2\sqrt{1+w^2/A^2}}\pvvhz\frac{1}{u^2}\mm {\rm{for}}\
          u> 0\ ,
       \nn\\
    &= &\lvvhz \frac{2mA}{w^2}
        -\frac{m}{A(1+w^2/A^2)^{3/2}}
        -\frac{2mA}{w^2\sqrt{1+w^2/A^2}}\pvvhz\frac{1}{u^2}\mm {\rm{for}}\
         u<0 \ ,
\EEA
with $w=\sn / u$ and thus the~function ${\cal K},_w(w)$ in (\ref{boost-c}) 
reads
\BEA
{\cal K},_w (w)&=&\frac{2mA}{w^2}
              +\frac{m}{A(1+w^2/A^2)^{3/2}}
              +\frac{2mA}{w^2\sqrt{1+w^2/A^2}} \mm {\rm{for}}\
           u> 0\  ,\nn\\
{\cal K},_w (w)&=&\frac{2mA}{w^2}
              -\frac{m}{A(1+w^2/A^2)^{3/2}}
              -\frac{2mA}{w^2\sqrt{1+w^2/A^2}} \mm {\rm{for}}\
          u<0\  .
\EEA
Following the~same procedure as in the~previous case, fairly long
calculations lead to the~total Bondi mass
(it is finite only for  $u<0$ where the~news function is regular)
\BEA
m&=&\frac{1}{8}
             \frac{m^2A^2\left |u\right |   \ln (A \left |u\right | ) (16+15A^2u^2)}
              { (A^2u^2+1)^{3/2}}
        -4m^2A^2 \left |u\right | \ln (A\left|u\right |)
          +2m^2A^2\left |u\right | \ln (A^2u^2+1)\nn\\
   &&\
          -\frac{1}{16}
                \frac{m^2A^2\left |u\right |\ln (A^2u^2+2+2\sqrt{A^2u^2+1})(16+15A^2u^2 )}
                      { (A^2u^2+1 )^{3/2}}
               -\frac{1}{8}\frac{m^2( A^2u^2+2 )}
         { (A^2u^2+1 )\left |u\right |}\ .
\EEA
In the~regular part of the~spacetime, for $u<0$,
the~dependence of  $m$ as a function of $u$
is qualitatively the~same as in the~previous case in Fig.~\ref{pichmotaBSzap}.
This solution  
 has the~axis regular only for $u<0$ where $c, _u/ \dsn$ is a  regular function  in
the~limit $\sn\ \msip\ 0$.

  \subsection{The~C-metric}

\subsubsection{Introduction}

The~C-metric was originally discovered in 1917 by Levi-Civita \cite{Levi} and Weyl \cite{Weyl}.
This algebraically degenerate vacuum solution of Petrov type D was called C-metric
by Ehlers and Kundt \cite{Ehlers} and its geometrical properties were studied in \cite{RT}.
Electromagnetic generalization of the~C-metric  was discovered in 1970 by Kinnersley and
Walker \cite{KWN}, who also first noticed that it includes not only static
regions when it  is analytically continued. 
 In the~early 80's  Ashtekar and Dray \cite{AshDN} showed
that the~C-metric is asymptotically flat at  spatial infinity and admits smooth cross-sections
of null infinity.
As the~original coordinate system corresponds to the~special algebraic  structure of the~solution,
it was necessary to transform the~C-metric into other coordinates where physical properties
can be examined (see \cite{KWN,FZN}). In 1983 Bonnor \cite{BonnorN} transformed the~C-metric 
into the~Weyl coordinates and further into coordinates adapted 
to boost and rotation symmetries (\ref{BStvar})
where also a radiative part of the~spacetime appears. Bonnor showed that the~C-metric could
be regarded as the~metric of two Schwarzschild black holes with  a string causing them to move 
with  a uniform acceleration. It is necessary to perform  long calculations to express
metric coefficients in terms of the~Weyl coordinates. In \cite{Bonnor2N} Bonnor showed
that the~charged C-metric with $m=0$ corresponds, in the~weak field limit, to the~electromagnetic Born solution
describing uniformly accelerated charges (see Subsec. \ref{secelmg}).
In 1995 Cornish and Uttley \cite{Cornish}
found that the~C-met\-ric in the~form (\ref{Cmetric}) contains four different spacetimes. 
Recently Yongcheng \cite{Yong} 
derived the~C-metric from the~metric of two superposed Schwarzschild
black holes assuming that 
the~mass and location of one of them approaches infinity in an appropriate way.
In 1998 Tomimatsu studied power-law decay of the~tails of gravitational radiation at future null infinity 
for the~C-metric and Wells \cite{WellsN} proved  the~black hole uniqueness theorem for it.

The~C-metric is a  solution of the~Einstein vacuum equations and in  coordinates $\{x$, $y$, $p$, $\t\}$
it  has the~form \cite{KWN}
\BE
{\rm d} s^2 = \frac{1}{A^2(x+y)^2} \left(F^{-1}{\rm d}y^2+G^{-1}{\rm d}x^2+
G {\rm d}p^2- F {\rm d} \tau^2  \right)  \ , \label{Cmetric}
\EE
where the~functions $F$, $G$ are cubic polynomials
\BEA
F&=&-1+y^2-2mAy^3 \ ,\\
G&=&\ \  1-x^2-2mAx^3 \ ,
\EEA
with  $m$ and $A$ being constant. We suppose that the~condition $27 m^2 A ^2 <1 $ is satisfied,
which implies that  $F$, $G$  have three distinct real roots $y_1$, $y_2$, 
$y_3$ and $x_1$, $x_2$, $x_3$.  

\subsubsection{Transformation of the~C-metric into the~Weyl form}

The~metric (\ref{Cmetric}) has two Killing vectors ($\partial /\partial \tau$ and $\partial /\partial p$) and it can be transformed into the~Weyl form  in specific regions.
Weyl solutions are static, axially symmetric spacetimes with the~metric
\BE
{\rm d}s^2={\rm e}^{-2U} \left[{\rm e}^{2 \nu}({\rm d}{\bar \rho}^2+{\rm d}{\bar z}^2)+
{\bar \rho}^2 {\rm d} \bar  \f^2 
\right] - {\rm e}^{2U} {\rm d} {\bar t}^2 \ , \label{Weylmetric}
\EE
where  the~functions $U({\bar \rho}, {\bar z})$ and $\nu({\bar \rho},{\bar z})$ are to be determined from
the~Einstein vacuum  equations
\BEA
0&=&\frac{\partial^2 U}{\partial {\bar \rho}^2} + \frac{1}{{\bar \rho}} \frac{\partial U}{\partial {\bar \rho}}
+ \frac{\partial^2 U }{\partial {\bar z}^2} \ , \nonumber \\
\frac{1}{{\bar \rho}} \frac{\partial \nu}{\partial {\bar \rho}} &=& \left(\frac{\partial U}{\partial {\bar \rho}}
 \right)^2 - \left(\frac{\partial U}{\partial {\bar z}} \right)^2  \ , \label{ErceW}\\
\frac{1}{{\bar \rho}} \frac{\partial \nu}{\partial {\bar z}}& =& 2 \frac{\partial U}{\partial {\bar \rho}}
\frac{\partial U}{\partial {\bar z}} \ . \nonumber
\EEA
The~first equation in (\ref{ErceW}) is the~integrability condition for the~other two ones.
If \mbox{the~$\bar z$-axis} given by $\bar \rho=0$ is regular, then for every infinitesimal circle around  it
the~ratio of circumference  to radius is $2\pi$ and thus the~Minkowski space exists locally.
The~regularity condition reads
\BE
{\rm e}^{2 \nu}(\bar \rho=0,\bar z) = 1 . \label{regcon}
\EE
A conical singularity appears on the~$\bar z-$axis if the~regularity condition (\ref{regcon}) 
 is not satisfied.

Using relations (see \cite{BonnorN}) ${\bar t}=\t$, ${\bar \varphi}=p$, and
\BEA
{\bar z}&=&\frac{1+mAxy(x-y)+xy}{A^2(x+y)^2} \ , \label{transfz}  \\
{\bar \rho}^2&=&\frac{FG}{A^4(x+y)^4} \label{transfrho} \ ,
\EEA
one can transform the~C-metric (\ref{Cmetric}) into the~Weyl form (\ref{Weylmetric}). 
It follows that the~metric functions ${\rm e}^{2U}$ and ${\rm e}^{2 \nu}$ are then given by
\BEA
{\rm e}^{2U}&=&\frac{F}{A^2(x+y)^2} \label{e2Uxy} \ , \\
{\rm e}^{2 \nu}&=& \frac{1}{A^4(x+y)^4\left[ \left(  \frac{\partial {\bar z}}{\partial y}\right)^2 +
\left( \frac{\partial {\bar \rho}}{\partial y}\right)^2 \right] }  \ . \label{e2nuxy}
\EEA
In order to express these functions in coordinates $\bar \rho$ and $\bar z$,  we have to invert
relations (\ref{transfz}) and (\ref{transfrho})  and substitute $x=x({\bar \rho},{\bar z})$ and $y=y({\bar \rho},{\bar z})$ into (\ref{e2Uxy}) and (\ref{e2nuxy}).

It is convenient to introduce functions $R_i$, $i=1,2,3$, defined by
\BE
R_i=\sqrt{({\bar z}-z_i)^2+{\bar \rho}^2} \ ,\label{RiCmetric}
\EE
where $z_1 < z_2 < z_3$ are the~roots of the~equation
\BE
2A^4 {z_i}^3 - A^2 {z_i}^2 + m^2 =0 \ .
\EE
The~functions  $R_i$ satisfy  relations \cite{BonnorN}
\BE
{R_i}^2=\left( \frac{m (A z_i)^{-1}+A^2 z_i (x-y)-mAxy }{A^2 (x+y)} \right)^2 \label{eqRn} \ ,
\EE
which, taking a square root,  become 
\BE
{R_i} =\epsilon_i \left( \frac{m (A z_i)^{-1}+A^2 z_i (x-y)-mAxy }{A^2 (x+y)} \right) 
\label{eqsqrRn}
\EE
with $\epsilon_1$, $\epsilon_2$, $\epsilon_3 = \pm 1$  to be chosen 
in such a way that the~right-hand
sides of (\ref{eqsqrRn}) are positive. Thus, all $\epsilon_i$ depend on $x$ and $y$.

Now one can express $x$ and $y$ in terms of $\bar \rho$, $\bar z$ 
\BEA
x &=& \frac{F_0+F_1}{2F_2} \ ,  \label{xrz} \\
y &=& \frac{F_0-F_1}{2F_2} \ , \label{yrz}
\EEA
where
\BEA
F_0 &=& \epsilon_1\epsilon_2\epsilon_3m(z_1-z_2)(z_2-z_3)(z_3-z_1)   \  , \nn\\
F_1 &=&- m\left[ \epsilon_1\epsilon_2(z_1-z_2)z_3R_3 +
\epsilon_2\epsilon_3(z_2-z_3)z_1R_1+\epsilon_3\epsilon_1(z_3-z_1)z_2R_2\right]  \ , \\
F_2&=&\frac{m^2}{2A}\left[ \epsilon_1\epsilon_2(z_1-z_2)R_3 +
\epsilon_2\epsilon_3(z_2-z_3)R_1
+\epsilon_3\epsilon_1(z_3-z_1)R_2\right] \ . \nonumber 
\EEA

Substituting the~equations (\ref{xrz}), (\ref{yrz}) into (\ref{e2Uxy}), 
(\ref{e2nuxy}), we get the~metric functions
${\rm e}^{2U}$ and ${\rm e}^{2 \nu}$ in terms of the~Weyl coordinates $\bar \rho$ and $\bar z$.
Since there are eight different choices for $\epsilon_1$, $\epsilon_2$, and $\epsilon_3$, there are
in principle eight different Weyl spacetimes contained in (\ref{Cmetric}). Four of them have signature
$+2$ and the~remaining four $-2$. 
Later we will consider some of these combinations explicitly.

\subsubsection{Properties of the~C-metric in the~original coordinates $x$, $y$ 
and in the~Weyl coordinates}

\begin{figure}
\begin{center}
\includegraphics*[height=6cm]{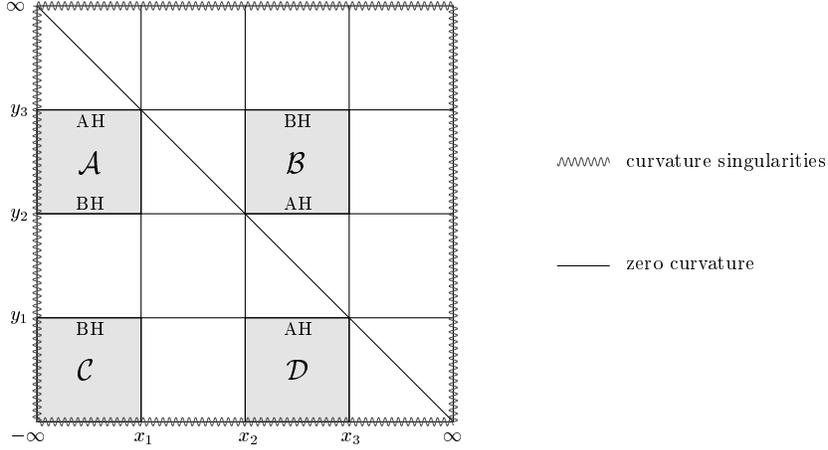}
\end{center}
\caption{The~character of the~regions determined by roots of polynomials $F$ and $G$ 
is schematically  illustrated.
Static regions are shaded, black hole horizons are denoted by
BH and acceleration horizons by AH. Curvature singularities and points with zero curvature
are also marked.}
\label{regpq}
\end{figure}

The~metric (\ref{Cmetric}) has signature $+2$ if $G>0$ and signature $-2$ if $G<0$. 
We choose signature $+2$ and thus we  investigate only regions where $G>0$. 
The~Killing vector $\partial /\partial \tau$=$\partial /\partial \bar t$ is timelike (and thus spacetime is static) for $F>0$. 
The~character of various regions is 
illustrated in Fig.~\ref{regpq}.  Notice that Fig. \ref{regpq} is
only schematic in the~sense that the~squares  have the~same size, although 
the~intervals between the~roots (or roots and infinity) are not the~same. 

As was mentioned in \cite{Cornish}, the~C-metric in the~$x$, $y$ coordinates contains
four various static regions with signature $+2$.   
These regions, denoted by ${\cal{A,\ B,\ C, \ D,}}$ 
have the~following values of  $\epsilon_i$:
\BEA
&&\epsilon_1=-1, \quad \epsilon_2=-1,  \quad  \ \epsilon_3=+1 \quad
{\mbox{in the~region }} \ {\cal A}, \nonumber \\
&&\epsilon_1=-1, \quad \epsilon_2=+1,  \quad  \ \epsilon_3=-1 \quad
{\mbox{in the~region }} \ {\cal B}, \\
&&\epsilon_1=+1, \quad \epsilon_2=+1,  \quad  \ \epsilon_3=+1 \quad
{\mbox{in the~region }} \ {\cal C}, \nonumber \\
&&\epsilon_1=+1, \quad \epsilon_2=-1,  \quad  \ \epsilon_3=-1 \quad
{\mbox{in the~region }} \ {\cal D}, \nonumber 
\EEA
and thus they can be transformed into different Weyl forms
by the~transformation (\ref{transfz}) and (\ref{transfrho}).

Now let us analyze singularities of the~metric (\ref{Cmetric}). For this purpose
it is useful to calculate several simple invariants of the~Riemann tensor :
\BEA
I_1 &=& {R^{\alpha \beta}}_{\gamma \delta} {R_{\alpha \beta}}^{\gamma \delta}
= -48 m^2 A^6 (x+y)^6 \ ,  \nonumber \\
I_2 &=& {R^{\alpha \beta}}_{\gamma \delta}  {R^{\gamma \delta}}_{\varepsilon \lambda}
{R^{\varepsilon \lambda}}_{\alpha\beta} = -96 m^3 A^9 (x+y)^9  \ , \label{Invars} \\
I_3 &=& R^{\alpha\beta\gamma\delta}R_{\alpha\varepsilon\lambda\mu}
{R^{\varepsilon\lambda}}_{\beta\nu} {R^{\mu\nu}}_{\gamma\delta} = -144 m^4A^{12} (x+y)^{12}\ .
\nonumber
\EEA      
A spacetime can be  flat only at those points where curvature invariants are approaching zero. 
Expressions
(\ref{Invars}) thus indicate that asymptotically  flat regions can  exist only on a  line given by 
\BE
x+y=0 \ .
\EE
Singularities of the~curvature invariants (\ref{Invars}) are located 
at the~points $x+y \rightarrow \infty$
and  marked in Fig. \ref{regpq}. 

From (\ref{transfrho}) it follows that if $F$ or $G$ vanishes, then the~Weyl coordinate
$\bar \rho$ is equal to zero.  Thus Eq. (\ref{transfrho})  maps the~boundaries  ($x=x_i$, $\pm \infty$) or 
 ($y=y_i$, $\pm \infty$) onto the~$\bar z-$axis of the~Weyl coordinates. Note that  as the~ Weyl coordinates 
describe only the~static part of the~spacetime there are regions of the~$\bar z$-axis   corresponding  to
black hole horizons or acceleration horizons. These segments of the~$\bar z$-axis do not correspond
to the~real, smooth physical  axis.

The~Killing vector $\der /\der\t$  
of the~metric (\ref{Cmetric})  
is timelike inside the~four regions ${\cal{A,\ B,\ C, \ D,}}$
but it has a zero norm at the~boundaries $y=y_i$  and thus the~Killing horizons are located there. To distinguish
between the~acceleration horizons and  black hole horizons, one  can calculate the~area of each horizon~\cite{Letelier}:
\BE
\int \limits_{0}^{2\pi} \int \limits_{x_k}^{x_{k+1}} \sqrt{g_{ \bar  \varphi  \bar  \varphi}} \sqrt{g_{xx}} 
{\rm d}x  {\rm d}  \bar  \varphi 
 = \frac{2 \pi}{A^2 } \int \limits_{x_k}^{x_{k+1}} \frac{{\rm d} x}{(x+y_j)^2}
=\frac{2 \pi}{A^2 } \frac{x_{k+1}-x_k}{(x_k+y_j)(x_{k+1}+y_j)} \label{Hsurf} \ ,
\EE 
where  $j=1,\dots, 3 $,  $k=0,1,\dots, 4$, $x_0=-\infty$, and  $x_4=\infty$.
A horizon of a black hole has a finite area  whereas an acceleration horizon  has an infinite area.
Locations of these horizons are indicated in Fig.~\ref{regpq}.

\begin{figure}
\begin{center}
\includegraphics{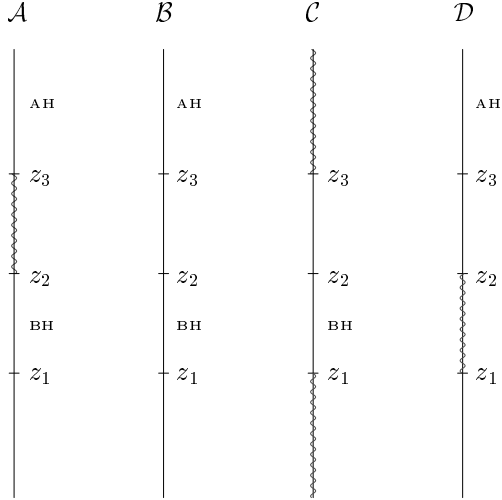}
\end{center}
\caption{The~$\bar z$-axis in the~Weyl coordinates for  cases ${\cal{A,\ B,\ C, \ D}}$.}
\label{figWeylaxes}
\end{figure}

Each region  ${\cal{A,\ B,\ C, \ D}}$ corresponds to a different Weyl metric 
(see  (\ref{e2UA})--(\ref{e2nuD})).   Properties of the~$\bar z$-axis in the~Weyl coordinates are illustrated
in Fig. \ref{figWeylaxes}.
 There is no horizon and  curvature singularity between
points  $z_2$ and $z_3$ in the~cases ${\cal{B,\ C, \ D}}$, however, in general, the~ axis is not regular there because of
a conical singularity (see Eq. (\ref{regcon})). 
The~same is true for ${\cal{A,\ B, \ D}}$ and $\bar z<z_1$.

There is no acceleration horizon in the~spacetime ${\cal{C}}$, and so this case 
 does not represent uniformly accelerated sources.
Moreover, this spacetime is not asymptotically flat because 
curvature invariants (\ref{Invars})  are everywhere non-vanishing and thus we will not study  this case further.

If functions ${\rm e}^{-2U}$ and ${\rm e}^{2 \nu}$ satisfy 
the~field equations (\ref{ErceW}), then the~same do also functions
\BEA
{\rm e}^{-2U'} &=& a\ {\rm e}^{-2U} \ , \label{newsol}\\
{\rm e}^{2\nu'} &=& b\  {\rm e}^{2\nu} \nonumber  \ 
\EEA
for $a$ and $b$ being constant.
This solution is for $a^2 \not= b^2$ indeed a new solution and cannot be transformed
into the~initial one by a coordinate transformation since
  a curvature invariant, which is coordinate independent, satisfies  
\BE
{R'}_{\alpha \beta \gamma \delta} {R'}^{\alpha \beta \gamma \delta} = \frac{a^2}{b^2}
{R}_{\alpha \beta \gamma \delta} {R}^{\alpha \beta \gamma \delta} \ .
\EE
Later we will use (\ref{newsol}) to  construct a solution which is regular for ${\bar z}<z_1$
or for $z_2<{\bar z}<z_3$ in the~cases where there is no curvature singularity.

Finally, let us express the~functions ${\rm e}^{2U}$ and ${\rm e}^{2\nu}$ in terms of ${\bar \rho}$,  ${\bar z}$
by substituting (\ref{xrz}), (\ref{yrz}) into (\ref{e2Uxy}), (\ref{e2nuxy}) 
(see  \cite{BonnorN} for the~case ${\cal B}$):
\BEA
{\rm e}^{2U}_{\cal A}&=&\frac{ \left[R_1-(\bar z-z_1) \right]  \left[R_2-(\bar z-z_2) \right]}
{R_3-(\bar z-z_3) } 
\quad {\mbox{in  the~region} }\  {\cal A},\label{e2UA}\\
{\rm e}^{2U}_{\cal B}&=&\frac{ \left[R_1-(\bar z-z_1) \right]  \left[R_3-(\bar z-z_3) \right]}
{R_2-(\bar z-z_2) } 
\quad {\mbox{in  the~region} }\  {\cal B},\\
{\rm e}^{2U}_{\cal D}&=&\frac{ \left[R_2-(\bar z-z_2) \right]  \left[R_3-(\bar z-z_3) \right]}{R_1-(\bar z-z_1) } 
\quad {\mbox{in  the~region} }\  {\cal D},
\EEA
and 
\BEA
{\rm e}^{2\nu}_{\cal A} &=& \frac{1}{4} \frac{m^2}{A^6 (z_3-z_1)^2 (z_3-z_2)^2} 
\frac{\left[R_2 R_3 +{\bar\r}^2 +({\bar z}-z_2)({\bar z}-z_3) \right]
      \left[R_3 R_1 + {\bar\r}^2 +({\bar z}-z_3)({\bar z}-z_1) \right]}
{R_1 R_2 R_3 \left[R_1 R_2 + {\bar\r}^2 + ({\bar z}-z_1)({\bar z}-z_2) \right]}\  {\rm e}^{2U}_{\cal A}\ , \\
{\rm e}^{2\nu}_{\cal B} &=& \frac{1}{4} \frac{m^2}{A^6 (z_2-z_1)^2 (z_3-z_2)^2} 
\frac{\left[R_2 R_3 + {\bar\r}^2 +({\bar z}-z_2)({\bar z}-z_3) \right]
      \left[R_1 R_2 + {\bar\r}^2 +({\bar z}-z_1)({\bar z}-z_2) \right]}
{R_1 R_2 R_3 \left[ R_1 R_3 + {\bar\r}^2 + ({\bar z}-z_1)({\bar z}-z_3) \right]}\  {\rm e}^{2U}_{\cal B}\ , \\
 {\rm e}^{2\nu}_{\cal D} &=& \frac{1}{4} \frac{m^2}{A^6 (z_1-z_2)^2 (z_1-z_3)^2} 
\frac{\left[ R_1 R_2 + {\bar\r}^2 +({\bar z}-z_1)({\bar z}-z_2) \right]
      \left[ R_1 R_3 + {\bar\r}^2 +({\bar z}-z_1)({\bar z}-z_3) \right]}
{R_1 R_2 R_3 \left[R_2 R_3 + {\bar\r}^2 + ({\bar z}-z_2)({\bar z}-z_3) \right]}\  {\rm e}^{2U}_{\cal D}  \ .
\label{e2nuD}
\EEA

\subsubsection{The~C-metric in the~canonical coordinates adapted to the~boost-rotation symmetry}

Following Bonnor \cite{BonnorN},  we shall transform the~metric from the~Weyl coordinates
to new coordinates in which the~boost-rotation symmetry of the~C-metric is evident. 
Performing an analytical continuation  of the~resulting metric,
two new regions of the~spacetime will appear.

The~metric of a  general boost-rotation symmetric spacetime  in polar coordinates $\{t$, $\rho$, $z$, $\phi\}$
has the~form (\ref{BStvar}) 
and the~transformation 
\BEA
\bar z -z_3 &=& \pul (t^2+\rho^2-z^2) \ ,\nonumber \\
{\bar \rho}^2 &=& \rho^2 (z^2-t^2) \ ,\label{trweylbs}\\
\bar t &=& {\mbox{arctanh}} (t/z)\ , \nonumber \\
\bar \phi &=& \phi \nonumber
\EEA
brings the~metric (\ref{Weylmetric}) into  this form with
\BE
{\rm e}^{\mu} = \frac{{\rm e}^{2U}}{z^2-t^2}  \  , \quad
{\rm e}^{\lambda} = 
\frac{{\rm e}^{2 \nu}}{{\rm e}^{2 U}} (\rho^2 + z^2 - t^2) \ .  \label{trel}
\EE
Applying the~transformation (\ref{trweylbs}), the~functions $R_1$, $R_2$, $R_3$ (\ref{RiCmetric})
turn to be
\BEA 
R_1 &=& \sqrt{(R+Z_1)^2-2Z_1\rho^2} \ , \nonumber \\
R_2 &=& \sqrt{(R+Z_2)^2-2Z_2\rho^2} \ , \\
R_3 &=& R \ , \nonumber
\EEA
where
\BEA
R &=& \pul (z^2-t^2+\rho^2) \ , \nonumber \\
Z_1 &=& z_1-z_3 \ , \\
Z_2 &=& z_2-z_3 \ . \nonumber
\EEA
The~functions ${\rm e}^{\mu}$ and ${\rm e}^{\lambda}$ are then given by
\BEA
\hspace{-1cm} {\rm {\rm e}}^{\mu}_{\cal{A}} &=& \frac{(R_1+R+Z_1-\rho^2)(R_2+R+Z_2-\rho^2)}{(z^2-t^2)^2} \ , \label{emuA}\\
\hspace{-1cm} {\rm {\rm e}}^{\mu}_{\cal{B}} &=& \frac{R_1+R+Z_1-\rho^2}{R_2+R+Z_2-\rho^2}  \ ,\label{emuB}\\
\hspace{-1cm} {\rm e}^{\mu}_{\cal{D}} &=& \frac{R_2+R+Z_2-\rho^2}{R_1+R+Z_1-\rho^2}  \ ,\label{emuD}
\EEA
and
\BEA
{\rm e}^{\lambda}_{\cal{A}} &=& \frac{1}{2} \frac{m^2}{A^6 {Z_1}^2 {Z_2}^2} 
\frac{\left[R(R_2+R+Z_2)-Z_2 \rho^2 \right]\left[R(R_1+R+Z_1)-Z_1 \rho^2 \right]}
{R_1R_2 \left[R_1 R_2 + (R+Z_1)(R+Z_2)-(Z_1+Z_2) \rho^2 \right]} \ ,\label{elA}\\
{\rm e}^{\lambda}_{\cal{B}} &=& \frac{1}{2} \frac{m^2}{A^6 {(Z_2-Z_1)}^2 {Z_2}^2} 
\frac{\left[R(R_2+R+Z_2)-Z_2 \rho^2 \right]\left[R_1R_2+(R+Z_1)(R+Z_2)-(Z_1+Z_2)\rho^2 \right]}
{R_1R_2 \left[R(R_1+R+Z_1) -Z_1 \rho^2 \right]} \ ,\label{elB}\\
{\rm e}^{\lambda}_{\cal{D}} &=& \frac{1}{2} \frac{m^2}{A^6 {(Z_2-Z_1)}^2 {Z_1}^2} 
\frac{\left[R(R_1+R+Z_1)-Z_1 \rho^2 \right]\left[ R_1R_2+(R+Z_1)(R+Z_2)-(Z_1+Z_2)\rho^2 \right]}
{R_1R_2 \left[R(R_2+R+Z_2) -Z_2 \rho^2 \right]} \ .\label{elD}
\EEA   

\begin{figure}
\begin{center}
\includegraphics*[height=5cm]{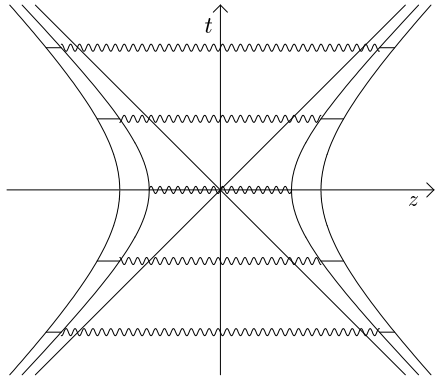}
\end{center}
\caption{The~type ${\cal A}$ solution for $ \alpha=1$, $\beta= 1/2$, $\gamma= \pm 1$  describes
two uniformly accelerated black holes connected by a curvature singularity.}
\label{hypAN}
\end{figure}

Let us recall the~regularity condition of the~metric (\ref{BStvar}) on the~roof 
(\ref{regstrecha}) 
\BE
{\rm e}^{\mu} = {\rm e}^{\lambda}     \mm  {\mbox{ for }} \mm    z^2=t^2 \ ,\label{regroof}
\EE
and the~regularity condition  of the~axis  (\ref{regaxis}) 
\BE    
\lim_{\rho \to 0} {\rm e}^{\lambda}  {\rm e}^{\mu} = 1 \ .
\EE
There are three regions on the~$z$-axis 
\BEA
{\mbox{ region I: }}\  && z^2-t^2 < -2 Z_2 \quad \dots  \quad {\mbox{between particles}},\nonumber \\
{\mbox{ region II: }}\  && -2Z_2<z^2-t^2 < -2 Z_1  \quad \dots  \quad  {\mbox{particles}}, \nonumber \\
{\mbox{ region III: }}\  && z^2-t^2 > -2 Z_1  \quad \dots  \quad  {\mbox{outside particles}}. \nonumber 
\EEA
Using the~multiplicative freedom described by Eq. (\ref{newsol})
for the~metric (\ref{BStvar})  with (\ref{emuA}), (\ref{elA}) in the~region ${\cal A}$, i.e., 
with 
\BEA
{\rm e}^{\mu} &=& \alpha \gamma \frac{(R_1+R+Z_1-\rho^2)(R_2+R+Z_2-\rho^2)}{(z^2-t^2)^2} \ , \label{emutA} \\
{\rm e}^{\lambda} &=& \beta \gamma
\frac{\left[ R(R_2+R+Z_2)-Z_2 \rho^2 \right]\left[R(R_1+R+Z_1)-Z_1 \rho^2 \right]}
{R_1R_2 \left[R_1 R_2 + (R+Z_1)(R+Z_2)-(Z_1+Z_2) \rho^2 \right]} \  , \label{eltA}
\EEA
we may construct solutions which are regular on the~whole roof and on a part of the~$z$-axis
(between or outside particles) as has been done in \cite{JiBiun,VPhD}.
This can be arranged by appropriately chosen  constants $\alpha$,  $\beta$,  $\gamma$.
The~regularity condition  on the~roof (\ref{regroof}) implies
\BE
\alpha = 2 \beta \quad {\mbox{and}} \mm Z_1<0\ ,\mm  Z_2 <0 \ .
\EE
To analyze the~regularity of the~axis, the~following 
limits of  functions ${\rm e}^{\mu}$ and ${\rm e}^{\lambda}$  for $\rho \to 0$ will be needed:
\BEA
\lim_{\rho \to 0}  {\rm e}^{\mu}& =&0\ ,\hspace{4cm}
\lim_{\rho \to 0}  {\rm e}^{\lambda} =0\hspace{4cm} {\mbox{ in  the~region I}},\nn\\
\lim_{\rho \to 0}  {\rm e}^{\mu}&=&0\ ,\hspace{4cm}
\lim_{\rho \to 0}  {\rm e}^{\lambda} 
          = -2\,{\frac {\beta\gamma {Z_1}^{2}\left (2 Z_2+  B \right )}
              {\left (2 Z_1 + B\right )\left ( Z_1- Z_2 \right )^{2}}}\mm
 {\mbox{ in  the~region II}},\nn\\
\lim_{\rho \to 0}  {\rm e}^{\mu} 
       &  =& {\frac {\alpha\gamma \left (2 Z_1+B \right )\left (2  Z_2+B \right )}{ B^2}}\ ,\mm
\lim_{\rho \to 0}  {\rm e}^{\lambda}
             = 2\,{\frac {{ B}^{2} \gamma \beta}{\left (2\,  Z_1 + B  \right )
               \left (2\,  Z_2 +  B \right )}}\mm \hspace{8pt}{\mbox{ in  the~region III}},   \nn
\EEA
where $B=z^2-t^2$ is defined in (\ref{KillMink}).
Thus the~only part of the~axis which can be regularized is  the~region III 
(outside black holes) with 
 \BE
 \alpha=1\ ,\mm \beta= \frac{1}{2}  \ ,\mm \gamma= \pm 1 \ ,
\EE
and then the~type ${\cal A}$ solution given by (\ref{emutA}),  (\ref{eltA}) describes two uniformly accelerated black holes
connected by a curvature singularity. The~rest of the~axis is regular (see Fig. \ref{hypAN}).

\begin{figure}
\begin{center}
\includegraphics*[height=5cm]{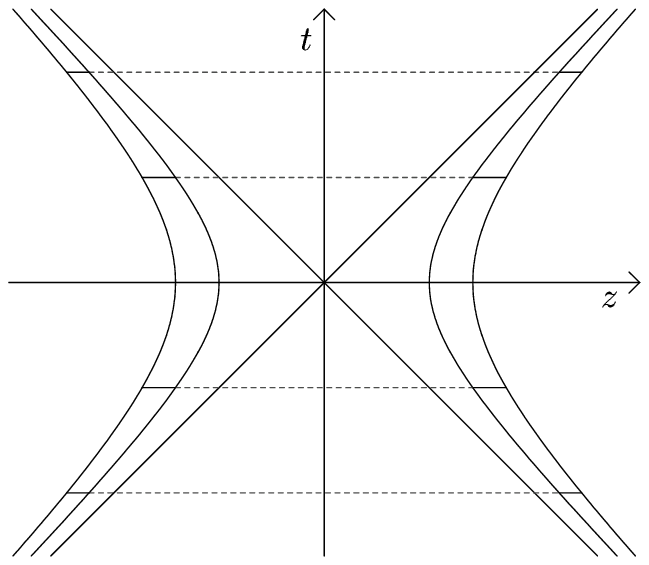}
\hspace{2cm}
\includegraphics*[height=5cm]{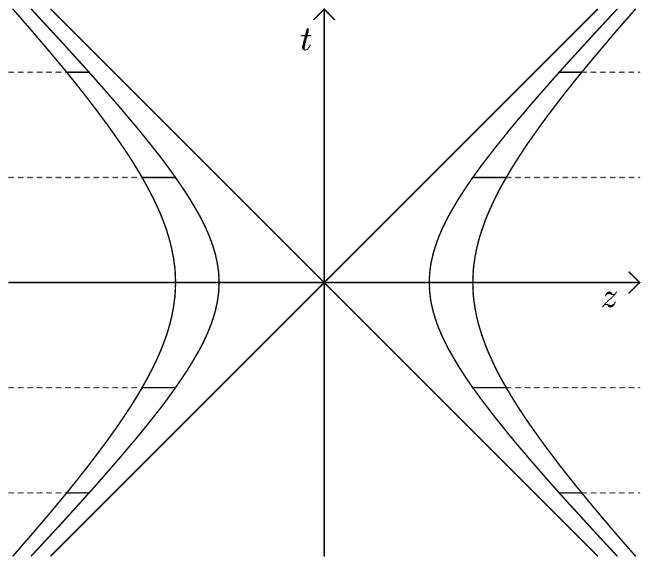}
\end{center}
\caption{
The~C-metric of the~type ${\cal B}$ with $\alpha=1$,  $\beta=1/2$, and \protect\\
a) with  $\gamma=\pm 1$ describes two uniformly accelerated black holes connected by a conical singularity;
\protect\\ 
b)  with $\gamma =\pm Z_1/Z_2$ describes  two uniformly accelerated black holes with conical
singularities extending  to infinity.}
\label{hypB}
\end{figure}

We use the~same procedure in the~case ${\cal B}$ (\ref{emuB}), (\ref{elB}):
\BEA
{\rm e}^{\mu} &=&\alpha \gamma  \frac{R_1+R+Z_1-\rho^2}{R_2+R+Z_2-\rho^2}   \ ,\nn \\
{\rm e}^{\lambda} &=& \beta \gamma
\frac{\left[ R(R_2+R+Z_2)-Z_2 \rho^2 \right]\left[R_1R_2+(R+Z_1)(R+Z_2)-(Z_1+Z_2)\rho^2 \right]}
{R_1R_2 \left[R(R_1+R+Z_1) -Z_1 \rho^2 \right]}  \  \label{emultB}
\EEA
to determine constants $\alpha$, $\beta$, and $\gamma$ 
regularizing the~roof and parts of the~axis. 
We find that for
\BE
\alpha=1\ , \mm  \beta=\frac{1}{2} \ ,\mm   \gamma=\pm 1\mm   
{\mbox{the~region III is regular}},
\EE
and the~C-metric of the~type ${\cal B}$  describes two black holes connected by a conical singularity, with the~rest of the~axis being regular
(see Fig. \ref{hypB}a).
For values
\BE
\alpha=1\ , \mm  \beta=\frac{1}{2} \ ,\mm \gamma =\pm \frac{Z_1}{Z_2} \mm  
{\mbox{the~region I is regular}},
\EE
and this case  describes  two uniformly accelerated black holes with
 conical singularities which are extending from black holes to infinity. 
The~axis is regular between
the~black holes (see Fig. \ref{hypB}b).

\begin{figure}
\begin{center}
\includegraphics*[height=5cm]{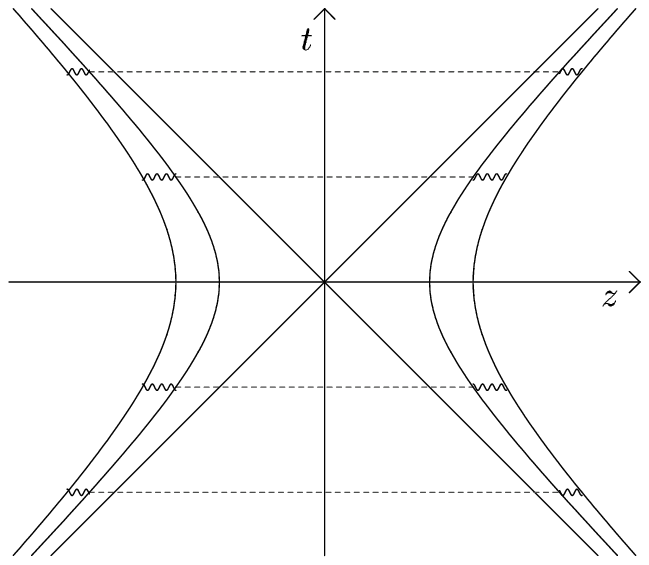}
\hspace{2cm}
\includegraphics*[height=5cm]{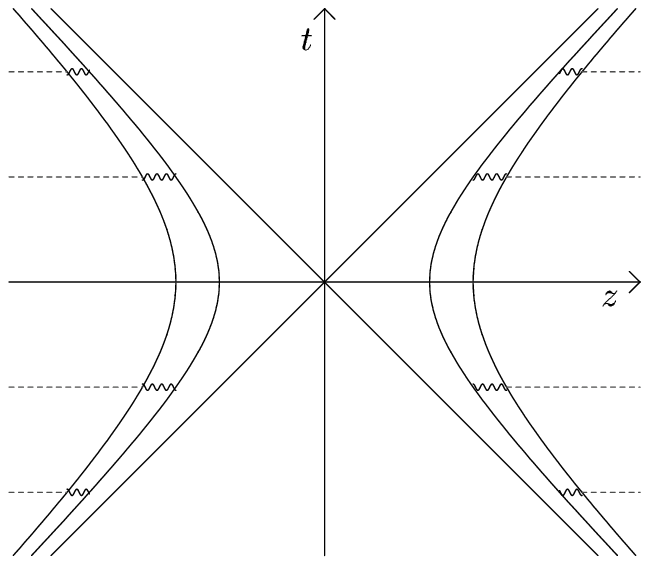}
\end{center}
\caption{ 
The~C-metric of the~type ${\cal D}$ with  $\alpha=1$, $\beta=1/2$, and\protect\\
a) with $\gamma = \pm 1$ corresponds to two uniformly accelerated  curvature singularities 
connected by a conical singularity;\protect\\
b) with  $\gamma = \pm Z_2/Z_1$ describes two uniformly accelerated curvature singularities with conical singularities which are extended  to infinity.}
\label{hypD}
\end{figure}

In the~${\cal D}$-case (\ref{emuD}), (\ref{elD}), i.e., 
for   functions ${\rm e}^{\mu}$ and ${\rm e}^{\lambda}$ being 
\BEA
{\rm e}^{\mu} &=&  \alpha \gamma \frac{R_2+R+Z_2-\rho^2}{R_1+R+Z_1-\rho^2} \nonumber \ , \\
{\rm e}^{\lambda}  &=& \beta \gamma
\frac{\left[R(R_1+R+Z_1)-Z_1 \rho^2 \right]\left[R_1R_2+(R+Z_1)(R+Z_2)-(Z_1+Z_2)\rho^2 \right]}
{R_1R_2 \left[R(R_2+R+Z_2) -Z_2 \rho^2 \right]}   \ , \nonumber
\EEA
the~axis is regular  in the~region III for values
\BE
\alpha=1, \ \beta=\frac{1}{2} , \ \gamma = \pm 1 \ .
\EE
Then  the~C-metric of the~type  ${\cal D}$ describes two
uniformly accelerated curvature singularities connected by a conical singularity. The~rest
of the~axis is regular (see Fig. \ref{hypD}a).
The~axis is regular in  the~region I for values
\BE
\alpha=1, \ \beta=\frac{1}{2} , \ \gamma = \pm \frac{Z_2}{Z_1} \ ,
\EE
that corresponds to two
uniformly accelerated curvature singularities with conical singularities extending  to infinity.
The~axis between curvature singularities is regular (see Fig. \ref{hypD}b).

\subsubsection{The~news function}

In further we calculate the~news function, $c,_{u}$,
 which characterizes gravitational radiation radiated out to null infinity (see Subsubsec.~\ref{secbondi}),
in the~most realistic case of the~C-metric, the~type ${\cal B}$. 
It is probably possible to express it explicitly  only in  a limit of small masses of the~sources. 
Introducing flat-space spherical coordinates
\BE
\rho = R \sin \vartheta \  ,\mm z=R \cos \vartheta  \ ,\mm t=U+R\ ,
\EE
and expanding  ${\rm e}^{\mu}$ and ${\rm e}^{\lambda}$ (\ref{emultB}) 
in powers of ${R}^{-1}$, with $U$, $\vartheta$, $\phi$ fixed, one gets
\BEA
{\rm e}^{\mu} &=& \alpha \gamma \left[ 1+\frac{1}{\sin^2 \vartheta } 
\left( {\sqrt{U^2-2Z_2 \sin^2 \vartheta }}-{\sqrt{U^2-2Z_1  \sin^2 \vartheta }} \right)
 \frac{1}{R} \right] + 
{\cal O} \left(R^{-2} \right) \label{Temr} \ ,\\
{\rm e}^{\lambda} &=& \beta \gamma \frac{\left[ U(U-U_2)-Z_2 \sin^2 \vartheta \right]
 \left[ U^2-(Z_1+Z_2) \sin^2 \vartheta+U_1 U_2 \right]}
{U_1 U_2 \left[U (U-U_1)-Z_1 \sin^2  \vartheta  \right]} + 
{\cal O} \left(R^{-1} \right) \ , \label{Telr} \ 
\EEA
where
\BE
U_1={\sqrt{U^2-2 Z_1 \sin^2 \vartheta }} \  , \mm
U_2={\sqrt{U^2-2 Z_2 \sin^2 \vartheta }} \  ,\nonumber 
\EE
and for  small $m$
\BE
Z_1=- \frac{1}{2A^2} - \frac{m}{A} + {\cal O}(m^2) \ ,\mm
Z_2=-  \frac{1}{2A^2} + \frac{m}{A} + {\cal O}(m^2) \ . 
\EE
Performing  an  expansion for small mass $m$  (with $A$ fixed), the~relations (\ref{Temr}) and (\ref{Telr})  yield
\BEA
{\rm e}^{\mu} &=& \alpha \gamma \left(1-\frac{2m}{\sqrt{A^2 U^2 + \sin^2 \vartheta }} \frac{1}{R} \right)  \label{Temk} \ ,\\
{\rm e}^{\lambda}  &=&  2 \gamma \beta - 8 \gamma \beta m A \sin^2 \vartheta
\frac{\sin^2 \ \vartheta - U A \sqrt{A^2 U^2 + \sin^2  \vartheta} + A^2 U^2 }{\left(A^2 U^2 + \sin^2  \vartheta \right) \left( 2 U^2 A^2 -2UA \sqrt{A^2 U^2 + \sin^2  \vartheta} + \sin^2 \vartheta \right)} 
 \ .  \label{Telk}
\EEA
Considering a conical singularity only between particles, i.e., putting
\mbox{$\alpha=1$}, \mbox{$\beta=1/2$}, \mbox{$\gamma=1$} and  following 
Sec.~4 in \cite{bicakBS}, we introduce functions $\bar \alpha (U,\vartheta)$ and $\bar B  (U,\vartheta)$ by
\BEA
{\rm e}^{\mu} &=& 1 + \frac{\bar \alpha}{R} + {\cal O} (R^{-2}) \label{defba} \ , \\
{\rm e}^{\lambda} &=& \bar B + {\cal O} (R^{-1}) \label{defB} \ ,
\EEA
where
\BEA
 \bar \alpha &=&  -\frac{2m}{\sqrt{A^2 U^2 + \sin^2 \vartheta }} \ ,\nn \\
\bar B  &=&  1 - 4  m A \sin^2 \vartheta  
\frac{\sin^2 \ \vartheta - U A \sqrt{A^2 U^2 + \sin^2  \vartheta} + A^2 U^2 }{\left(A^2 U^2 + \sin^2  \vartheta \right) \left( 2 U^2 A^2 -2UA \sqrt{A^2 U^2 + \sin^2  \vartheta} + \sin^2 \vartheta \right)} \ .
\EEA

In order to analyze radiative properties of the~system, we move to the~Bondi  coordinates $u$, $r$,
$\theta$, $\phi$  (see Subsubsec.~\ref{secbondi}) connected by an asymptotic transformation (see \cite{bicakBS})
\BEA
U &=& \pi^0 (u,\theta) + {\cal O} \left(r^{-1}\right) \nonumber \ ,\\
R &=& r  + {\cal O} ({r}^{0}) \nonumber \ ,\\
\vartheta &=& \theta  + {\cal O} \left(r^{-1}\right) \ ,\nonumber
\EEA
where
$\int {\bar B}(\pi^0 ,\theta) {\mbox d} \pi^0 =  u $, 
and for small $m$  one gets 
$ \pi^0 =  u$ and thus $U = u$. 
To express the~news function we use  the~relation (26) of Ref.~\cite{bicakBS} 
\BE
c,_{u} = \frac{1-\bar B+{\bar \alpha},_{U} \sin^2 \vartheta}{2 \bar B \sin^2 \vartheta}
\EE
which is valid for a general boost-rotation symmetric spacetime that  is asymptotically
flat, 
does not contain  an infinite cosmic string, and has hypersurface orthogonal Killing vectors
as was shown in \cite{BicTN}.
Then the~news function for the~ ${\cal B}$-type  C-metric regularized outside the~particles
is, in terms of $w=\sn / u$, given by  \cite{JiBiun,VPhD}
\BEA
c,_{u} &=& {\frac {-mA\left (A-2\,\sqrt {{A}^{2}+{w}^{2}}\right )w^2}{{u}^{2}
\left ({A}^{2}+{w}^{2}\right )^{3/2}\left (2\,{A}^{2}-2\,\sqrt {{A}^{2
}+{w}^{2}}A+{w}^{2}\right )}} \quad {\mbox{for $u > 0$}} \label{cup}\  ,\\
c,_{u} &=& {\frac {mA \left (A+2\,\sqrt {{A}^{2}+{w}^{2}}\right )w^2}{{u}^{2}
\left ({A}^{2}+{w}^{2}\right )^{3/2}\left (2\,{A}^{2}+2\,\sqrt {{A}^{2
}+{w}^{2}}A+{w}^{2}\right )}} \quad {\mbox{for $u < 0$}} \ . \label{cum}  
\EEA
This is in agreement with the~expression  (\ref{boost-c}) obtained in \cite{ajajibi} for a general boost rotation symmetric spacetime.
The~news function  is regular for $u<0$ and the~total Bondi mass calculated there
has qualitatively the~same dependence on $u$ as in the~BS-solutions (see Subsec.~\ref{secBS} and 
Fig.~\ref{pichmotaBSzap}a therein).
The~news function is singular for $u>0$ at $\theta=0, \ \pi$
due to a conical  singularity connecting the~particles (recall that we assume $A>0$).

  \subsection {The~spinning C-metric}

The~spinning C-metric (SC-metric), a generalization of the~C-metric  with the~rotation, NUT parameter, 
and electric and magnetic charges, was found by Pleba\' nski and Demia\' nski \cite{PlebDem}
in 1976  in the~coordinates $\{\tau$, $p$, $q$, $\sigma\}$ 
\BEA
{{\rm d}s}^{2}=
{\frac {E\left (Q-P{q}^{4}\right )}{F}} \ {{{\rm d}\t}}^{2}
-\ {\frac {EF}{Q}}\ {{\rm d}q}^{2}
-\ {\frac {EF}{P}}\ {{\rm d}p}^{2} 
+\ {\frac {E\left (Q{p}^{4}-P\right )}{F}}\ {{\rm d}\s}^{2}
-\ {\frac {2\, E\left (Q{p}^{2}+P{q}^{2}\right )}{F}}\  {\rm d}\t {\rm d}\s\  ,
\label{RotC}
\EEA
where
\BEA
P&=&\ \ \g-{ \varepsilon}\,{p}^{2}+2\,m{p}^{3}-\g{p}^{4} \label{eqP} \ , \\
Q&=&-\g+{ \varepsilon}\,{q}^{2}+2\,m{q}^{3}+\g{q}^{4} \label{eqQ} \ ,  \\
E&=&\left (p+q\right )^{-2} \ , \label{eqE} \\
F&=&1+{p}^{2}{q}^{2} \ , \label{eqF}
\EEA
$m$, $\g$, $\varepsilon$ being constant. Here  we put the~parameters 
corresponding to the~NUT parameter, electric and magnetic charges, 
and cosmological constant, $n$, $e$, $g$, $\lambda$ respectively,  equal to zero,
however,  in the~general form of the~Pleba\'nski-Demia\'nski metric (Eqs. (2.1), (3.25) in \cite{PlebDem})
they are non-vanishing.
They also demonstrated that the~standard C-metric (\ref{Cmetric}) or
the~Kerr metric can be obtained 
from (\ref{RotC}) by specific limiting procedures. 
This metric was later analyzed by Farhoosh and Zimmerman  \cite{FZN}, 
and recently discussed by Letelier and Oliveira \cite{Letelier}. 

Hereafter we assume
the~polynomials $P$ and $Q$ to have four distinct
real roots (the~corresponding conditions for $m$, $\varepsilon$ and $\g$ are given in
\cite{bivoj}). 

The~metric (\ref{RotC}) having two Killing vectors,
\BE
{\frac{\partial}{\partial \tau}}  \quad {\rm and} \quad  {\frac{\partial}{\partial \sigma}} \ , \label{Killpq}
\EE
was transformed into coordinates adapted to boost-rotation symmetry in \cite{bivoj}.
However, the~corresponding Killing vectors are not hypersurface orthogonal and thus the~general theory
summarized in Subsec.~\ref{secgen} cannot be applied to this case and until now no similar 
general theory   is available 
for the~boost-rotation symmetric  spacetimes with Killing vectors which are not hypersurface orthogonal.

\subsubsection{The~SC-metric in the~Weyl-Papapetrou coordinates}

As was shown in \cite{bivoj} the~SC-metric (\ref{RotC}) can be converted 
by a transformation 
\BEA
{ {\rm d}\t}&=&{ \kappa_1}\,{ {\rm d}\bar t}+{ \kappa_2}\,{ {\rm d}\bar \phi} \ , \nonumber \\
{ {\rm d}\s}&=&{ \kappa_3}\,{ {\rm d}\bar t}+{ \kappa_4}\,{ {\rm d}\bar \phi}  \label{eqtrKill}
\EEA
with  constants $ \kappa_1 \dots  \kappa_4$ and
\BEA
{\bar \rho}^2&=&{E}^{2}{\cal K}^{2}PQ \ ,     
\ \ \  {\cal K}  =\left ({ \kappa_2}\,{\kappa_3}-{ \kappa_1}\,{ \kappa_4}\right ) \ ,\label{tranWrho}\\
\bar z& =& {\cal K} {\frac { -\gamma-\varepsilon\,qp-m{q}^{2}p+mq{p}^{2}+\gamma{q}^{2}{p}^{2}
 }{ (p+q )^{2}}} \  \label{tranz} 
\EEA
into the~standard Weyl-Papapetrou
form
\BE
{\rm d}s^2 ={\rm e}^{-2U} \left [ {{\rm e}^{2 \nu}} ({{\rm d} \bar \rho}^{2}
+{{ {\rm d} \bar z}}^{2} )+{\bar \rho}^{2}{{{\rm d}\bar \phi}}^{2} \right ] -{{\rm e}^{2\,U}}
\left ({ {\rm d} \bar t}+a{ {\rm d} \bar \phi}\right )^{2}  \ ,   
\label{WeylPap}
\EE
where  $\bar \phi  \in \langle 0, 2 \pi)$,  $\bar \rho \in \langle 0,\infty)$, $\bar t,\bar z \in R$
and functions $U$,  $\nu$, $a$ depend only on $\bar \rho$ and $\bar z$.
The~metric functions satisfy the~vacuum Einstein equations 
\BEA
0&=& \frac{\partial^2 U}{\partial \bar z^2} + \frac{1}{\bar \rho} \frac{\partial U}{\partial \bar \rho}
+ \frac{\partial^2 U}{\partial \bar \rho^2} +\frac{1}{2} \frac{{\rm e}^{4U}}{\bar \rho^2}
\left [ \left (\frac{\partial a}{\partial \bar \rho} \right )^2 +
\left (\frac{\partial a}{\partial \bar z}  \right )^2 \right ] \ , \nonumber \\
0&=& \frac{\partial }{\partial \bar z} \left ( \frac{{\rm e}^{4U}}{{\bar \rho}}
\frac{\partial a}{\partial \bar z} \right ) +
\frac{\partial }{\partial {\bar \rho}} \left ( \frac{{\rm e}^{4U}}{{\bar \rho}}
\frac{\partial a}{\partial {\bar \rho}} \right )  \ ,  \\
\frac{1}{{\bar \rho}} \frac{\partial \nu}{\partial {\bar \rho}}& =&
\left (\frac{\partial U}{\partial {\bar \rho}} \right )^2
-\left (\frac{\partial U}{\partial \bar z} \right )^2 -\frac{{\rm e}^{4U}}{4 {\bar \rho}^2}
\left [ \left (\frac{\partial a}{\partial {\bar \rho}} \right )^2 -
\left (\frac{\partial a}{\partial \bar z}  \right )^2 \right ] \ , \nn\\
\frac{1}{{\bar \rho}} \frac{\partial \nu}{\partial \bar z}& =& 2 \frac{\partial U}{\partial {\bar \rho}}
\frac{\partial U}{\partial \bar z} - \frac{{\rm e}^{4U}}{2 {\bar \rho}^2}
\frac{\partial a}{\partial {\bar \rho}} \frac{\partial a}{\partial \bar z} \ . \nonumber
\EEA
Notice   that analogically to the~non-rotating case,  the~$\bar z$-axis of the~Weyl-Papapetrou
coordinates (${\bar \rho} = 0$) is a real smooth geometrical (physical) axis only at points  where the~metric (\ref{WeylPap}) satisfies the~regularity conditions
\BE
\lim_{{\bar \rho} \to \infty} \frac{X_{,\alpha} X^{,\alpha}}{4X} = 1\ , \  {\rm where}\  X =\xi_{(\bar \phi) \alpha} \xi_{(\bar \phi) }^\alpha  \ , \mm \xi_{(\bar \phi)} = \frac{\partial}{\partial \bar \phi}\  .   \label{regosa}
\EE
Some parts of the~axis 
may represent either a horizon of a rotating black hole or a rotating string and thus the~regularity conditions
(\ref{regosa}) are not satisfied. 

\subsubsection{Properties of the~SC-metric in the~
$p$, $q$ coordinates}

In \cite{bivoj} 
the~character of the~Killing vectors (\ref{Killpq}) and their linear combinations with constant
coefficients $k_1$, $k_2$, 
\BE
\eta = k_1 \frac{\partial}{\partial \tau}  + k_2  \frac{\partial}{\partial \sigma} \label{combkill} \ ,
\EE
is studied. 
There exists a linear combination (\ref{combkill}) that is spacelike and another one which is timelike
in regions where 
\BE
PQ>0 \ ,\label{statcond}
\EE
and thus the~spacetime is  stationary there.

The~metric (\ref{RotC}) has the~signature $-2$ for $P>0$, whereas $P<0$ implies signature $+2$.
Following \cite{bivoj} we choose the~signature $+2$, i.e.,  $P<0$.

The~curvature invariant
\BEA
I &=& R^{\alpha\beta\gamma\delta}R_{\alpha\varepsilon\lambda\mu}
{R^{\varepsilon\lambda}}_{\beta\nu} {R^{\mu\nu}}_{\gamma\delta} =
-144m^4 \frac{(p+q)^{12}}{(1+p^2q^2)^6} \ \label{invar}
\EEA
indicates that asymptotically flat regions can  exist only on the~line given by 
\mbox{$p+q=0$} and the~curvature singularities are located at points 
 $[p,q]=[0,\infty]$,
$[0,-\infty]$, $[-\infty,0]$,  $[\infty,0]$, 
where the~invariant (\ref{invar}) diverges.

\begin{figure}
\begin{center}
\includegraphics*[height=6cm]{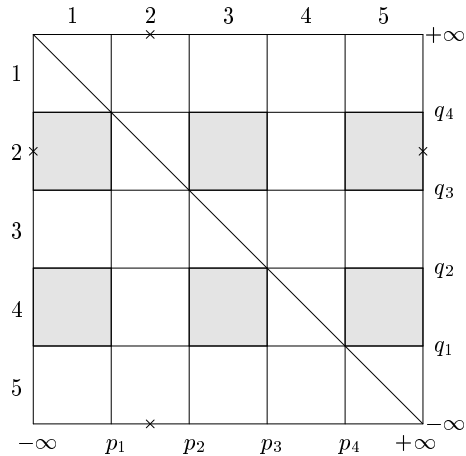}
\end{center}
\caption{The~schematic illustration of the~SC-metric
in the~$p$, $q$ coordinates. } 
\label{picscpqsq}
\end{figure}

These results are summarized in Fig.~\ref{picscpqsq}, where  
$p_1 < p_2 < p_3 < p_4$ and $q_1 < q_2 < q_3 < q_4$
denote the~roots of the~polynomials $P$ and $Q$, and the~numbers $1,\  \dots, \ 5$ indicate regions
between individual roots or between a root and infinity. Stationary regions with the~signature $+2$
are shaded, the~diagonal line corresponds to the~asymptotically flat regions and each curvature singularity
is indicated by a cross. A detailed analysis (see \cite{bivoj}) shows that the~left edge of the~picture
has to be identified with the~right one and  the~bottom edge with the~upper one as well.
Notice that lines $p=p_i$ or $q=q_i$  are mapped onto the~$\bar z$-axis of the~Weyl coordinates
due to the~relation (\ref{tranWrho}).

\begin{figure}
\begin{center}
\includegraphics*[height=45mm]{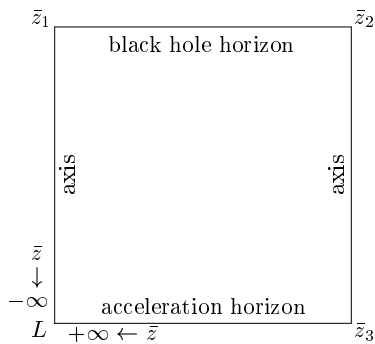}
\nopagebreak
\caption{The~square $\{2,3\}$ from Fig.~\ref{picscpqsq}.}
\label{picmalyctvr}
\end{center}
\end{figure}

Further we concentrate on the~square $\{2,3\}$ with $p \in \langle p_2 , p_3 \rangle$,  $q \in \langle q_3, q_4 \rangle$ as in \cite{bivoj}. The~detailed analysis in \cite{bivoj} determines the~location of the~black
hole horizon, the~acceleration horizon, and the~geometrical axis  on \mbox{the~${\bar z}$-axis} of the~Weyl coordinates
along which  the~string represented by a conical singularity may lay (see Fig.~\ref{picmalyctvr}). The~vortices are marked by ${\bar z}_i$, the~values of which follow from (\ref{tranz}).
The~lower left vortex $L$ has a special property: when it is approached along the~left edge 
of the~square we arrive at ${\bar z}\msip - \infty$, along the~bottom edge at ${\bar z}\msip + \infty$ 
and by approaching it along different  lines from the~interior of the~square, we can achieve various values
of ${\bar \r}$ and  ${\bar z}$.

\subsubsection{The~SC-metric in the~canonical coordinates adapted to the~boost-rotation symmetry}

The~Weyl-Papapetrou coordinates cover only stationary regions of the~SC-metric (shadowed squares in Fig.~\ref{picscpqsq}). 
The~~transformation  (\ref{trweylbs}) brings the~metric (\ref{WeylPap}) into the~form
 in which the~boost-rotation symmetry of the~SC-metric will become manifest
\BEA
{\rm ds}^2  &=&  {\rm e}^{\lambda} {\rm d} \rho^2 + \rho^2 {\rm e}^{-\mu} {\rm d} \phi^2 + 
  \frac{1}{z^2-t^2} \left[ ({\rm e}^{\lambda} z^2 - {\rm e}^{\mu} t^2 ) {\rm d} z^2 
-   2zt ({\rm e}^{\lambda} - {\rm e}^{\mu}  ) {\rm d} z  {\rm d} t 
+     ({\rm e}^{\lambda} t^2 - {\rm e}^{\mu} z^2 ) {\rm d} t^2  \right]      \nonumber \\ 
 &&  -   2a {\rm e}^{\mu} (z {\rm d}t -  t {\rm d} z)  {\rm d} \phi -a^2 {\rm e}^{\mu} (z^2-t^2)   {\rm d} \phi^2 \ ,
\label{BStvarR}
\EEA
where  ${\rm e}^\m$ and ${\rm e}^\l$ are given by (\ref{trel}).
As in the~non-rotating case, by analytically continuing the~resulting metric, two new radiative regions
of the~spacetime will arise, however,   parts of the~spacetime under the~black hole horizons, 
which are not included in the~Weyl-Papapetrou form, are neither involved here.

The~metric (\ref{BStvarR}) is a generalization of the~metric (\ref{BStvar}) that can be obtained by
putting $a=0$ in (\ref{BStvarR}). 
The~coefficients $\k_1$,  $\k_2$, $\k_3$, $\k_4$  in the~relation (\ref{eqtrKill})  are chosen to  make the~metric
smooth across the~roof,  null cone and on the~axis $\rho=0$ except the~part of the~axis between black 
holes, where a nodal singularity occurs causing the~acceleration, 
and also to make the~metric asymptotically Minkowskian at spatial infinity. 
These constants  are  expressed in  terms of $m$, $\varepsilon$, and $\g$ in \cite{bivoj}.

There are two qualitatively new features of the~SC-metric in the~form (\ref{BStvarR})
in comparison to the~non-rotating C-metric: there exist two symmetrically located 
small regions bellow the~roof
near the~black hole horizons where the~boost Killing vector,
\BE
\frac{\der}{\der {\bar t}}=z\frac{\der}{\der t}+t\frac{\der}{\der z}\ ,
\EE
is spacelike which correspond to ergoregions; 
there also exist two regions in the~vicinity with the~edges of the~nodal singularity where causality violation, 
$g_{\phi \phi} < 0$, occurs.

\begin{figure}
\begin{center}
\includegraphics*[height=5.5cm]{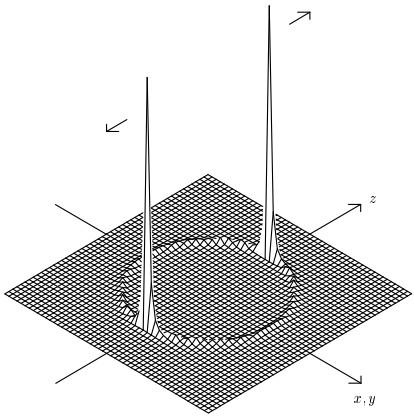}
\end{center}
\caption{The~curvature invariant  (\ref{invar}) at a fixed time $t=t_0>0$;
the~peaks correspond to the~black hole horizons. 
The~gravitational radiation pulse  propagates in all directions  with the~velocity of light
and a  decreasing amplitude.}
\label{picpulsinv}
\end{figure}

Radiative properties of this spacetime has not yet been  analyzed rigorously, however, the~invariant
(\ref{invar}) at a fixed time (Fig.~\ref{picpulsinv}) having the~character of a pulse
demonstrates the~radiative character of the~SC-metric. This pulse character of the~radiation,
first noticed in \cite{bicakBS}, is a typical feature  of 
boost-rotation symmetric solutions. 

\begin{center}
{\bf{ACKNOWLEDGMENTS}}\\
\end{center}
We are grateful to  prof.~J.~{Bi\v c\' ak} for introducing us into the topic and 
supervising our doctoral theses.


\end{document}